\documentclass[twocolumn,showpacs,preprintnumbers,amsmath,amssymbaps,pra,citeautoscript]{revtex4}

\usepackage{graphicx}
\usepackage{amssymb}
\usepackage{amsfonts}
\usepackage{dcolumn}
\usepackage{bm}
\usepackage{float}
\usepackage{footnote}
\usepackage{cases}
\usepackage{amsmath}
\usepackage{color}
\usepackage[font=small,labelfont=bf]{caption}

\newcommand{\ket}[1]{\left| #1 \right>} 
\newcommand{\bra}[1]{\left< #1 \right|} 

\newcommand{\req}[1]{Eq.~(\ref{#1})}

\newcommand{\lrangle}[1]{\left< #1\right>}

\numberwithin{equation}{section}

\begin{document}


\title{Solitons in One Dimensional Systems at BCS-BEC Crossover}

\author{Tianhao Ren}
 \email{tr2401@columbia.edu}
\author{Igor Aleiner}
 \email{aleiner@phys.columbia.edu}
\affiliation{Physics Department, Columbia University, New York, NY 10027, USA}

\date{\today}

\begin{abstract}
We developed a comprehensive semiclassical theory of solitons in one dimensional systems at BCS-BEC crossover to provide a semiclassical explanation of their excitation spectra. Our semiclassical results agree well with the exact solutions on both the deep BCS and deep BEC side and explain qualitatively the smooth crossover between them. Especially, we showed that the minimum energy of the $S=1/2$ excitation is achieved exactly at the Fermi momentum $k_F=\pi n/2$, where $nm_F$ ($m_F$ is the mass of the fermionic atom) is the total mass density of the system. This momentum remains unchanged along the whole crossover, whether the mass is contained in the bosonic molecules as on the deep BEC side or in the fermionic atoms as on the deep BCS side. This phenomenon comes about as a result of a special feature of one dimensional systems that the conventional quasiparticle is not stable with respect to soliton formation. It is valid not only in exactly solvable models but also on the level of semiclassical theory. Besides, we also resolved the inconsistency of existing semiclassical theory with the exact solution of soliton-like $S=0$ excitations on the deep BCS side by a new proposal of soliton configuration.
\end{abstract}


\maketitle

\section{Introduction}
Soliton formation is an important and rich nonlinear phenomenon in various branches of physics. In many exactly solvable models, both classical and quantum mechanical ones, soliton plays a unique role. It is well known that the interacting bosons in one dimension (the Lieb-Liniger model) show an unexpected branch in its excitation spectrum, usually referred to as the type-II excitations \cite{PhysRev.130.1605,PhysRev.130.1616}. Later it was found that the interacting fermions in one dimension (the Yang-Gaudin model) have a similar phenomenon \cite{RevModPhys.85.1633,1367-2630-18-7-075004}. The fact that they originate from solitons can be clearly seen in the semiclassical analysis, where solitons serve as an alternative solution to the semiclassical equation of motion apart from the spatially homogeneous solution \cite{Kulish:1976ek,PhysRevA.91.023616}.

It is even more interesting, as we will show, that these soliton-like solutions can further affect the spin excitations in a striking way that they will fix the minimum energy of the spin excitations exactly at momentum $k_F=\pi n/2$, where $nm_F$ ($m_F$ is the mass of the fermionic atom) is the conserved total mass density of the system and it remains unchanged along the whole crossover.

This is in sharp contrast to the situation in higher dimensions, whereby tuning interaction along the BCS-BEC crossover we can move this momentum from $k_F$ on the deep BCS side to zero on the deep BEC side \cite{2015qgee.book..179P}. In this paper, we present a comprehensive semiclassical theory of solitons in one dimensional systems at BCS-BEC crossover, where we explain the soliton interpretation of the type-II excitations and the fixing of the momentum for the minimum energy of spin excitations. Our theory explains the semiclassical origin of the excitation spectrum of the Yang-Gaudin model, where existing semiclassical proposals fail to reconcile with the exact solutions \cite{PhysRevA.91.023616,1367-2630-18-7-075004}. Our theory also serves as yet another example of the dramatic effect solitons can have on low dimensional physics.

In the next section, we will review the exact solutions of the Lieb-Linger model, the Yang-Gaudin model and the model of BCS-BEC crossover in one dimension. From there, we raise the questions mentioned above and we further analyze them in the sections to follow. We first outline the general formalism of the semiclassical analysis in presence of solitons across the BCS-BEC crossover. We then apply it to the $S=1/2$ and $S=0$ excitations respectively, where we present analytic analysis on both deep BCS and deep BEC side and qualitative analysis for the crossover. Finally, we summarize the main results and make the conclusion.

\section{Review of Exact Solutions and their Relation to Solitons}
The model of interacting bosons and fermions in one dimension can be both solved exactly via the technique of Bethe ansatz \cite{Korepin_1993}, the former is known as the Lieb-Liniger model \cite{PhysRev.130.1605,PhysRev.130.1616}, and the latter is known as the Yang-Gaudin model \cite{GAUDIN196755,PhysRevLett.19.1312}. An exactly solvable model connecting them to describe BCS-BEC crossover in one dimension can also be constructed \cite{PhysRevLett.93.090408,PhysRevLett.93.090405,Ren}. In this section, we present the excitation spectra of these exactly solvable models. In the $S=0$ excitations (where $S$ is the total spin) for all these models, there is an extra soliton-like branch apart from the usual Bogoliubov quasiparticle branch. In the $S=1/2$ excitations, one finds the minimum of the energy lying exactly at the Fermi momentum $k_F=\pi n/2$. These are the key features we would like to explain when later developing the corresponding semiclassical theory.

We start with the Lieb-Liniger model, described by the Hamiltonian
\begin{equation}
\label{eq:HLL}
\hat{\mathcal{H}}=\int dx\left[\partial_x\hat{\varphi}^{\dagger}(x)\partial_x\hat{\varphi}(x)+c_B\hat{\varphi}^{\dagger}(x)\hat{\varphi}^{\dagger}(x)\hat{\varphi}(x)\hat{\varphi}(x) \right],
\end{equation}
where $\hat{\varphi}$ represents the spinless bosons with mass $m_B=1/2$, and $c_B>0$ corresponds to the repulsion between bosons. Also we adopt the convention that $\hbar=1$ in this paper.

\begin{figure}
\includegraphics[scale=0.5]{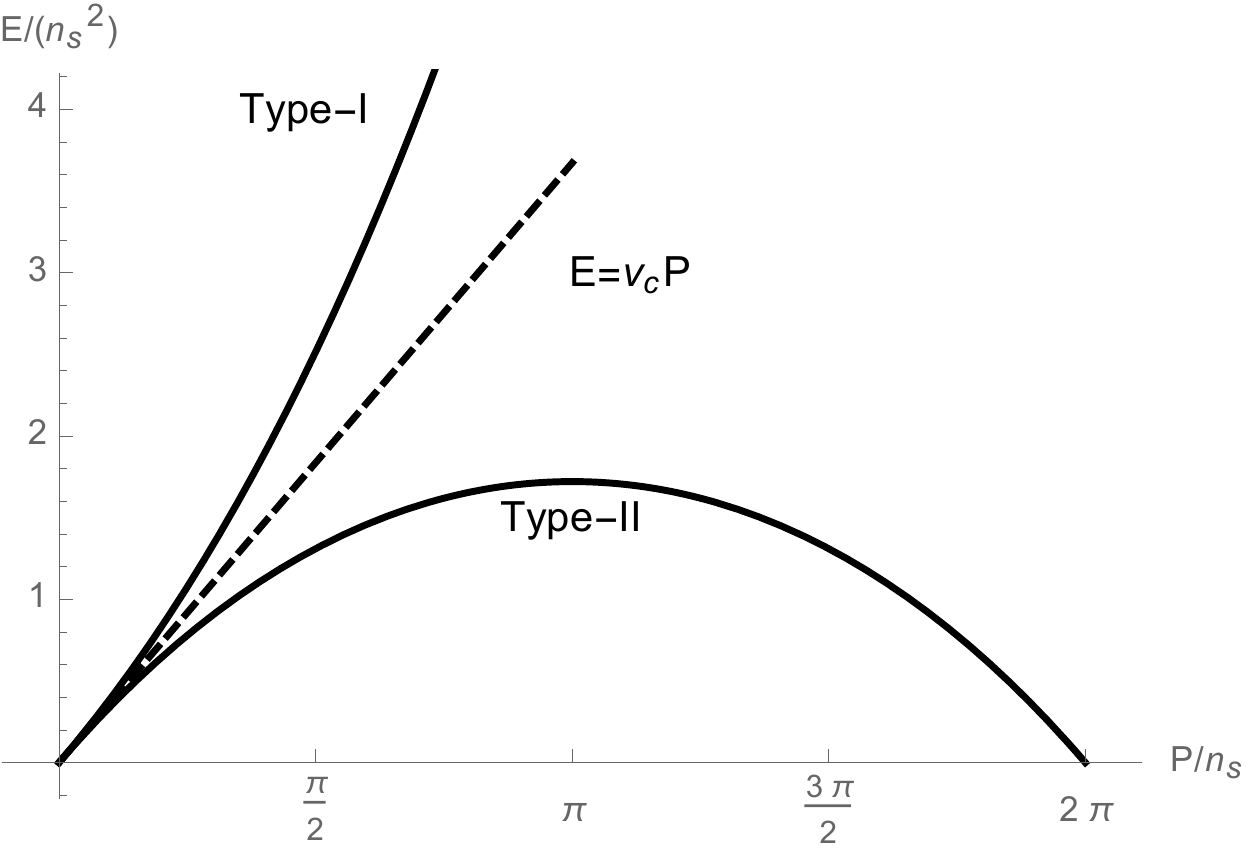}
\caption{\footnotesize The typical excitation spectrum of the Lieb-Liniger model, calculated for coupling strength $\gamma=c_B/n_s=0.43$. There are two branches, type-I for Bogoliubov quasiparticles and type-II for soliton-like excitations. Also shown in the figure is the sound velocity $v_c$, which scale as $\sqrt{c_Bn}$.}
\label{fig:LL}
\end{figure}

A typical excitation spectrum of Lieb-Liniger model is shown in Fig. \ref{fig:LL}.
It is composed of two branches, the usual Bogoliubov quasiparticle (Lieb-Liniger type-I) branch, and the Lieb-Liniger type-II branch. At long wavelength, both branches reduce to a linear dispersion as phonons, with the same sound velocity $v_c=\sqrt{c_Bn}$, whose magnitude decreases with the coupling strength. The key features of the type-II excitations are that it has $\epsilon(2\pi n_s) \to 0$ as the system size goes to infinity, $L\to\infty$, and it has its maximum energy achieved at momentum $k=\pi n_s$. This periodicity of the type-II branch is a consequence of translational invariance, where the shift of momentum for each boson by the amount of $2\pi/L$ costs $(n_sL)(2\pi/L)^2\to 0$ in energy but changes the total momentum by $(n_sL)(2\pi/L)=2\pi n_s$ \cite{Ren}. Similarly, the total energy also remains invariant under the momentum reflection $k\to 2\pi/L-k$ for each boson, which means the spectrum has an additional symmetry of reflection about total momentum $\pi n_s$. As a result, the maximum of the spectrum is fixed at momentum $\pi n_s$. It is known that this point corresponds to a motionless (dark) soliton, and all the Lieb-Liniger type-II branch has the physical interpretation as the dispersion relation $E(P)$ for the moving (grey) soliton with velocity $v_s=\partial E(P)/\partial P$ \cite{Kulish:1976ek,PhysRevA.78.053630}.

\begin{figure}
\includegraphics[scale=0.5]{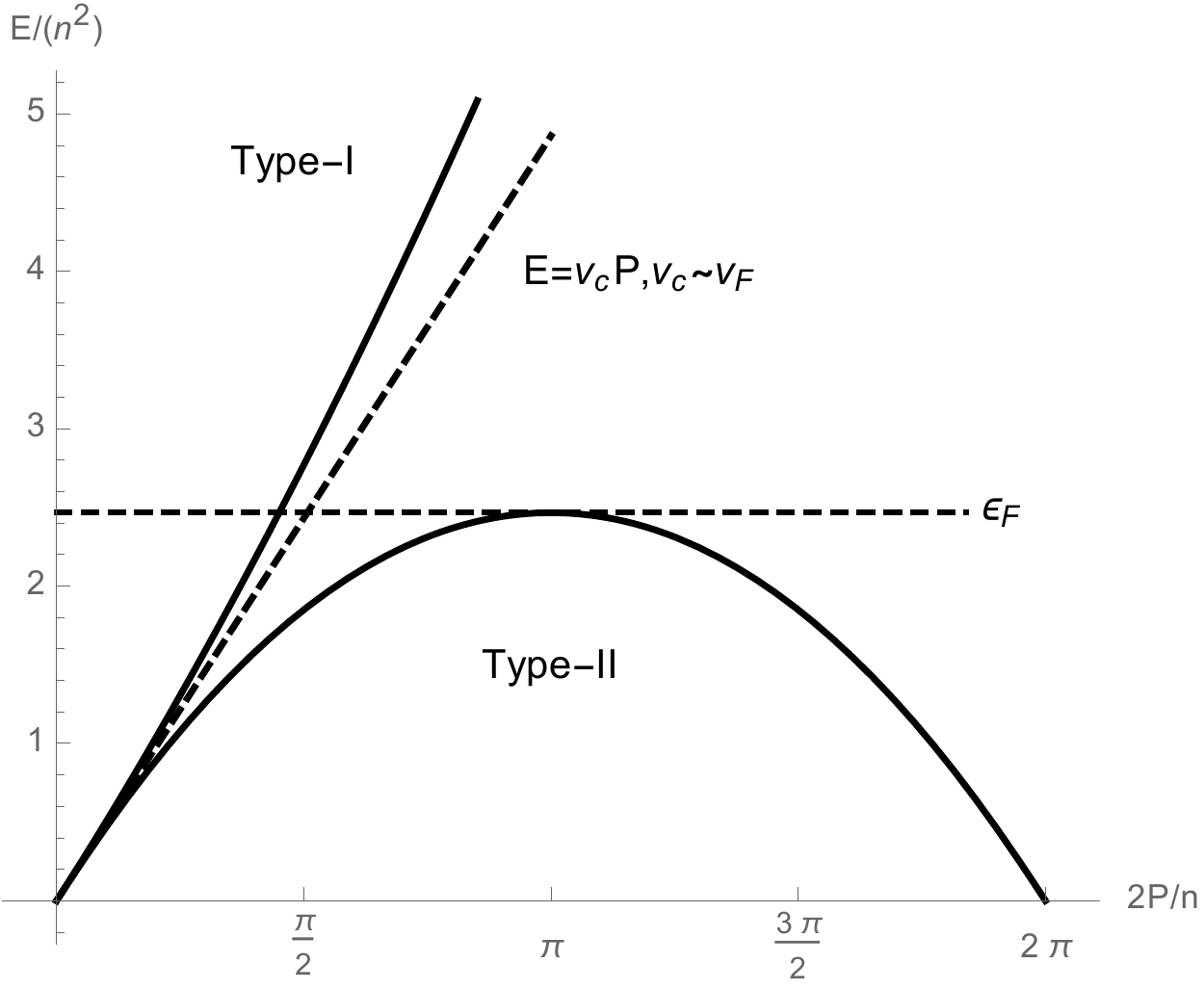}
\caption{\footnotesize The typical $S=0$ excitation spectrum of the Yang-Gaudin model, calculated for coupling strength $\gamma=c_F/n=0.15$. There are also two branches, type-I for Bogoliubov quasiparticles and type-II for soliton-like excitations. Also shown in the figure is the sound velocity and the Fermi energy $\epsilon_F$, we can see in the weak coupling limit, the dark soliton has an energy on the scale of $\epsilon_F$ and the sound velocity is on the scale of $v_F$.}
\label{fig:YG1}
\end{figure}

Now we move on to the attractive Yang-Gaudin model, which is defined by the following Hamiltonian:
\begin{equation}
\label{eq:HYG}
\hat{\mathcal{H}}=\int dx\left[\partial_x\hat{\psi}^{\dagger}(x)\partial_x\hat{\psi}(x)-c_F\hat{\psi}^{\dagger}(x)\hat{\psi}^{\dagger}(x)\hat{\psi}(x)\hat{\psi}(x) \right],
\end{equation}
where $\hat{\psi}=\begin{pmatrix}\hat{\psi}_{\uparrow}\\ \hat{\psi}_{\downarrow}\end{pmatrix}$ represents the $S=1/2$ fermions with mass $m_F=1/2$, and $c_F>0$ corresponds to the attraction between fermions. This attraction, however weak, produces bound pairs in one dimension. A typical spectrum of $S=0$ excitations of the Yang-Gaudin model is shown in Fig. \ref{fig:YG1},
which is pretty similar to the one we obtain in the Lieb-Liniger model. The notable differences here are the scale of the maximum energy of type-II excitations and the sound velocity. In the weak coupling limit $c_F/n \ll 1$, the maximum energy is on the scale of the Fermi energy $\epsilon_F=\pi^2n^2/4$ and the sound velocity is on the scale of the Fermi velocity $v_F=\pi n$. Since the velocity is large when $k\to 0$, there is no semiclassical description for the dispersion relation, but near the maximum of the spectrum where the velocity is small, a semiclassical description is still possible. The recent attempt by ~\citet{PhysRevA.91.023616} to develop such a description led to incorrect energy scale and curvature near the maximum of the spectrum \cite{1367-2630-18-7-075004}. We are going to reconcile this discrepancy in this paper.

 In the strong coupling limit $c_F/n \gg 1$ where the fermions are tightly bounded, instead of behaving like a system of weakly coupled bosons, the Yang-Gaudin model produces a system of hardcore bosons know as the fermionic super Tonks-Girardeau gas \cite{RevModPhys.85.1633}. As a result, the sound velocity is still on the scale of the Fermi velocity, and the spectrum of Fig. \ref{fig:YG1} preserves qualitative shape for any value of $c_F$.

\begin{figure}
\includegraphics[scale=0.5]{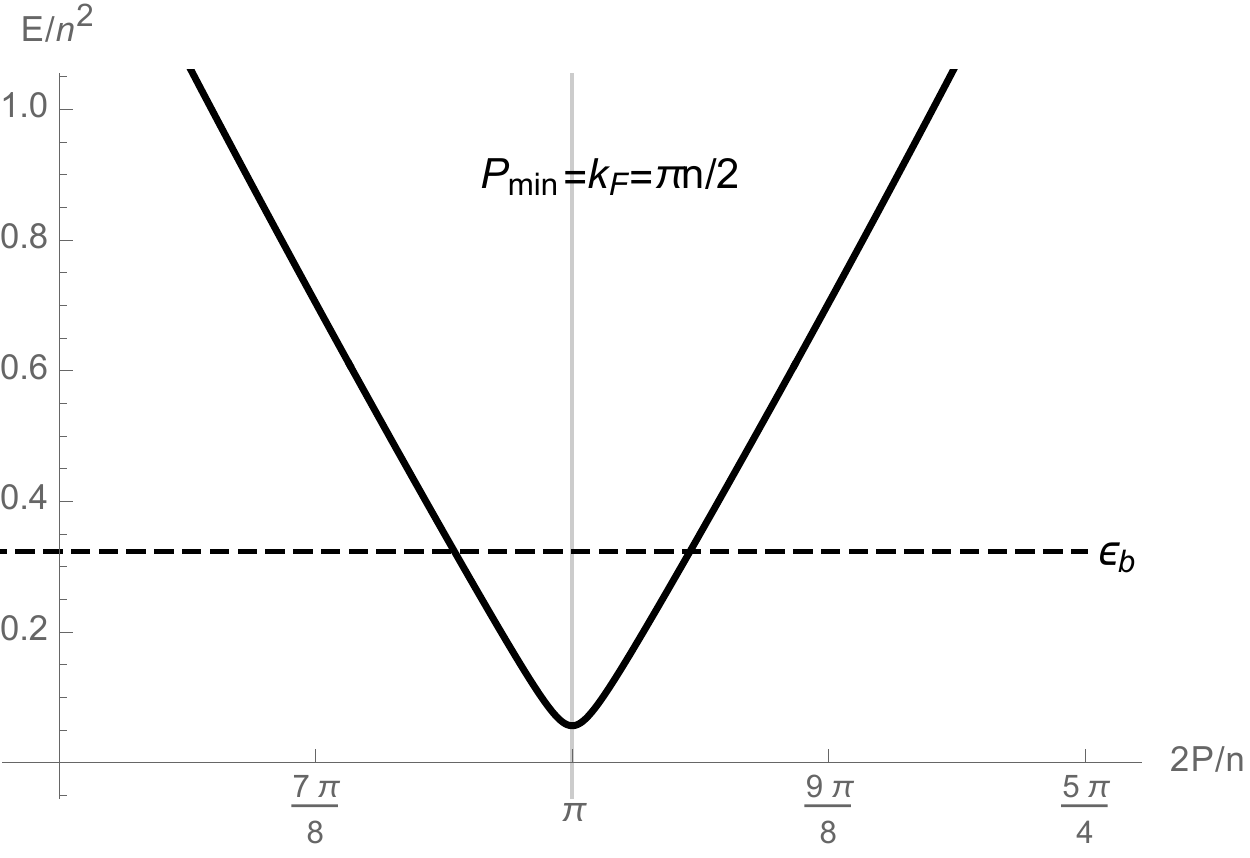}
\caption{\footnotesize The typical $S=1/2$ excitation spectrum of the Yang-Gaudin model, calculated for coupling strength $\gamma=c_F/n=1.13$. The minimum energy is obtained at the Fermi momentum $k_F=\pi n/2$, with a small region of quadratic spectrum around it. Also shown in the figure is the binding energy $\epsilon_b$ for the singlet pairs, which is bigger than the spin gap.}
\label{fig:YG2}
\end{figure}

\begin{figure}
\includegraphics[scale=0.5]{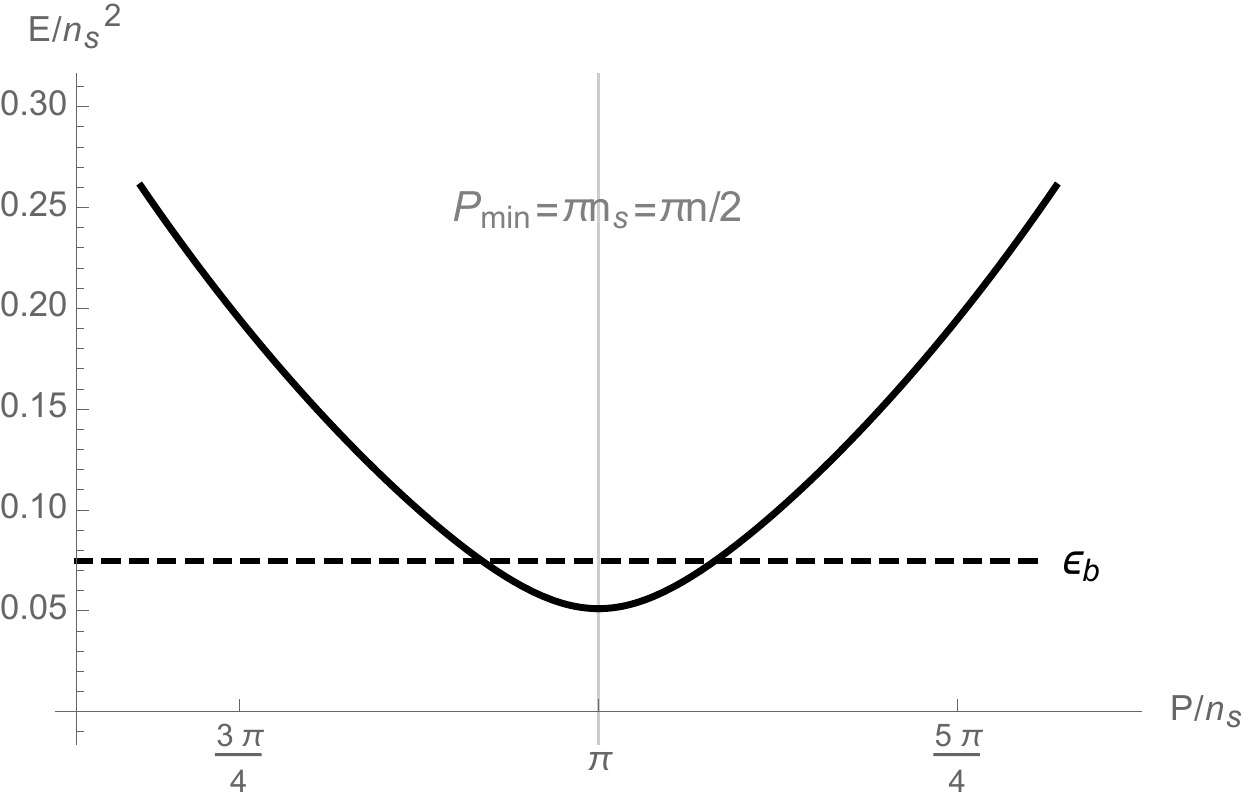}
\caption{\footnotesize The typical $S=1/2$ excitation spectrum on the BEC side, calculated for coupling strength $\gamma_1=c_1/n=0.34$ and $\gamma_2=c_2/n=0.27$. In the plotting scale we have used $n_s=n/2$. The minimum energy is again obtained at the Fermi momentum $k_F=\pi n/2$. Also shown in the figure is the binding energy $\epsilon_b$ for the singlet pair, which is bigger than the spin gap.}
\label{fig:BEC}
\end{figure}

A typical spectrum of $S=1/2$ excitations of the Yang-Gaudin model is shown in Fig. \ref{fig:YG2},
where the minimum energy is achieved exactly at the Fermi momentum $k_F=\pi n/2$, irrespective of the coupling strength. This exactness is unusual, since it is without the correction on the scale of $\delta k\sim \Delta_0/v_F$ that would be introduced by the conventional BCS theory in the weak coupling limit (where $\Delta_0$ is the gap width), and it is contrary to the usual conclusion that the minimum energy should be achieved at zero momentum in deep BEC regime in higher dimensions \cite{2015qgee.book..179P}. At first sight, this could be caused by the fact that the strong coupling limit $c_F/n\gg 1$ of Yang-Gaudin model is not a system of weakly coupled bosons, which invalidates it as a proper model for BCS-BEC crossover. To test this idea, we recently proposed a new model of BCS-BEC crossover subject to exact solutions by Bethe ansatz \cite{Ren}. The fermionic version of this model is described by the Hamiltonian:
\begin{equation}
\label{eq:fermionicmodel}
\begin{split}
\hat{\mathcal{H}}= & \int dx \Big{\{}  \partial_x\hat{\psi}^{\dagger}\partial_x \hat{\psi}+\frac{1}{2}\partial_x\hat{\bm{a}}^{\dagger}\cdot\partial_x\hat{\bm{a}} +\frac{1}{2}\partial_x\hat{b}^{\dagger}\partial_x\hat{b} \\
& -\epsilon_a\hat{\bm{a}}^{\dagger}\cdot\hat{\bm{a}}-\epsilon_b\hat{b}^{\dagger}\hat{b}+\lambda_{\psi}\hat{\psi}^{\dagger}\hat{\psi}^{\dagger}\hat{\psi}\hat{\psi}\\
&+\left[\frac{t_a}{2}\left(i\partial_x\hat{\psi}^T\bm{\sigma}\sigma_y\hat{\psi}\right)\cdot\hat{\bm{a}}^{\dagger}+h.c.\right]\\
&+\left[\frac{t_b}{2}\left(i\hat{\psi}^T\sigma_y\hat{\psi}\right)\cdot\hat{b}^{\dagger}+h.c.\right] \Big{\}},
\end{split}
\end{equation}
where $\bm{\sigma}=(\sigma_x,\sigma_y,\sigma_z)$ is the Pauli matrix, $\hat{\psi}=\begin{pmatrix}\hat{\psi}_{\uparrow}\\ \hat{\psi}_{\downarrow}\end{pmatrix}$ represents the fermions with mass $m_F=1/2$ and $\lambda_{\psi}$ is the repulsive coupling between them. $\hat{\bm{a}}$ represents the vector resonance at energy $-\epsilon_a$ and $\hat{b}$ represents the scalar resonance at energy $-\epsilon_b$, both of which are of mass $m_a=m_b=1$. Both of the resonances are needed for the exact solvability, which can be achieved by fine tuning the position of the resonant levels. The behavior of this model is then controlled by two parameters:
\begin{equation}
c_1=|t_a|^2/4, ~~~ c_2=c_1+|t_b|^2/(2\epsilon_b).
\end{equation}
This model has the Lieb-Liniger model and the Yang-Gaudin model as its two limits in the parameter range $c_1\sim c_2$ and $c_1\gg c_2$ respectively, thus providing a model of BCS-BEC crossover in one dimension that is subject to exact solutions. On the side where it reduces to the Yang-Gaudin model with $c_F=c_2$, the excitation spectrum is basically the same as shown in Fig. \ref{fig:YG1} and Fig. \ref{fig:YG2}; On the side where it reduces to the Lieb-Liniger model with $c_B=c_1-c_2$, the $S=0$ spectrum is basically the same as shown in Fig. \ref{fig:LL}. In addition to that,  we also have $S=1/2$ excitations now, whose typical behavior is shown in Fig. \ref{fig:BEC}.
We can see that the spectrum has the same feature as that on the BCS side, with the minimum energy still obtained exactly at the Fermi momentum $k_F=\pi n/2$. even though the $S=0$ sector corresponds to weakly interacting bosons with $v_c\ll v_F$.

In all the exactly solvable models presented above, the fixing of minimum spin excitation energy at $k_F$ is a phenomenon robust against variations of coupling constants across the whole range, which is in sharp contrast to the situation in higher dimensions \cite{2015qgee.book..179P}. It leads us to the conclusion that this is most probably a general feature not limited to exact solvability. One may suspect that the fixing is a consequence of the Luttinger theorem, but this is not true due to the fact that the system here is gapped and there is no conservation of the number of fermions (since there is tunneling between atoms and molecules back and forth). On the other hand, the maximum of the $S=0$ excitations can be interpreted as a dark soliton, with the spectrum near it as a moving grey soliton. We propose that the minimum of the $S=1/2$ excitations is also a dark soliton with one extra fermion bounded on it and $k_F$ is just the momentum of this dark soliton, whereas the fermion sitting bounded on top of it doesn't bring any new momentum. This will be done in the next sections.

\section{General Formalism}
For the purpose of semiclassical analysis, let's consider the following simplified model of BCS-BEC crossover at the mean field level:
\begin{equation}
\label{eq:BCS-BEC}
\begin{split}
\hat{\mathcal{H}}=& \int dx~\left\{\partial_x\hat{\psi}^{\dagger}\partial_x\hat{\psi}+\frac{1}{2}\partial_x\hat{b}^{\dagger}\partial_x\hat{b}-\epsilon_b\hat{b}^{\dagger}\hat{b} \right.\\
&\left.+\left[\frac{t_b}{2}\left(i\hat{\psi}^T\sigma_y\hat{\psi} \right)\hat{b}^{\dagger}+h.c. \right]\right\}-\mu\hat{\mathcal{N}},\\
\hat{\mathcal{N}}=&\int dx \left(\hat{\psi}^{\dagger}\hat{\psi}+2\hat{b}^{\dagger}\hat{b}\right),\\
\hat{\mathcal{P}}=&~\frac{1}{2i}\int dx\left( \hat{\psi}^{\dagger}\partial_x\hat{\psi}+\hat{b}^{\dagger}\partial_x\hat{b}-h.c. \right),
\end{split}
\end{equation}
where $\hat{\psi}=\begin{pmatrix}\hat{\psi}_{\uparrow}\\ \hat{\psi}_{\downarrow}\end{pmatrix}$ represents the $S=1/2$ fermions with mass $m_F=1/2$, $\hat{b}$ with mass $m_b=1$ represents a scalar resonance with resonant energy $-\epsilon_b$ when $\epsilon_b<0$ or a molecule with binding energy $\epsilon_b$ when $\epsilon_b>0$. The coupling constant $t_b$ is chosen to be real. Operator $\hat{\mathcal{N}}$ is a conserved quantity and the expectation value of $m_F\hat{\mathcal{N}}=\hat{\mathcal{N}}/2$ gives out the total mass of the system. Although not subject to exact solutions, this model grasps the essence of the BCS-BEC crossover and is more friendly to semiclassical analysis.

A conventional way to analyze the semiclassical origin of the excitations is to treat the operators as classical fields and to solve the semiclassical equations of motion for them. Its validity can be justified via the saddle point approximation in the path integral formalism. The symmetry-broken ground state of the system is then represented by the expectation value $\lrangle{\hat{b}}=b_0$, where $b_0$ is a constant, and the excitations are represented by a space-time varying expectation value $b(x,t)\equiv \lrangle{\hat{b}}$, where we use the periodic boundary condition such that $b(x,t)=b(x+L,t)$. As we treat the operator $\hat{b}$ as a classical field $b(x,t)$, the part of $\hat{\mathcal{H}}$ that involves fermionic operators can be diagonalized via the Bogoliubov-Valatin transformation
\begin{equation}
\begin{pmatrix}\hat{\psi}_{\uparrow}\\ \hat{\psi}^{\dagger}_{\downarrow}\end{pmatrix}=\sum_n\begin{pmatrix}u_n(x,t) & -v^*_n(x,t)\\ v_n(x,t) & u^*_n(x,t)\end{pmatrix}\begin{pmatrix}\hat{\gamma}_{n\uparrow} \\ \hat{\gamma}^{\dagger}_{n\downarrow}\end{pmatrix}
\end{equation}
to the following Hamiltonian
\begin{equation}
\label{eq:meanpsi}
\hat{\mathcal{H}}_{\psi}=\kern-0.5em\sum_{\epsilon_n>0}\left[-\epsilon_n(\Delta,\Delta^*)+\epsilon_n\Big{|}_{t_b=0} \right]+\kern-0.5em\sum_{\epsilon_n>0,\sigma}\kern-0.5em\epsilon_n\hat{\gamma}^{\dagger}_{n\sigma}\hat{\gamma}_{n\sigma},
\end{equation}
where we have defined $\Delta(x,t)\equiv t_bb(x,t)$ and the classical fields $u_n(x,t), v_n(x,t)$ satisfy the Bogoliubov-de Gennes equation \cite{Schrieffer_1983} with periodic boundary conditions:
\begin{equation}
\label{eq:eqforuv}
\begin{split}
	& \begin{pmatrix}
		-\partial^2_x-\mu & \Delta \\ \Delta^* & \partial^2_x+\mu
	  \end{pmatrix}\begin{pmatrix}
		u_n \\ v_n
	  \end{pmatrix}=\epsilon_n\begin{pmatrix}
		u_n \\ v_n
	  \end{pmatrix}, \\
	& ~~\begin{cases}u_n(x+L,t)=u_n(x,t)\\v_n(x+L,t)=v_n(x,t)\end{cases}.
\end{split}
\end{equation}
Using these classical fields $b(x,t)$, $u_n(x,t)$ and $v_n(x,t)$, the energy and momentum of the system under a particular filling configuration of \req{eq:meanpsi} can then be expressed as
\begin{equation}
\label{eq:exforEP1}
\begin{split}
	& E=\int dx\left( \frac{1}{2}|\partial_xb|^2-(2\mu+\epsilon_b)|b|^2 \right)+E_{\psi},\\
	&E_{\psi}=\kern-0.5em\sum_{\epsilon_n>0}\left[-\epsilon_n(\Delta,\Delta^*)+\epsilon_n\Big{|}_{t_b=0} \right]+\kern-0.5em\sum_{\epsilon_n>0,\sigma}\kern-0.5em\epsilon_n\lrangle{\hat{\gamma}^{\dagger}_{n\sigma}\hat{\gamma}_{n\sigma}},\\
	&P= \int dx\left(\sum_{\epsilon_n>0}\frac{u^*_n\overleftrightarrow{\partial_x}u_n+v^*_n\overleftrightarrow{\partial_x}v_n}{2i}\sum_{\sigma}\lrangle{\hat{\gamma}^{\dagger}_{n\sigma}\hat{\gamma}_{n\sigma}} \right)\\
	&~~~+\int dx\left(\sum_{\epsilon_n>0}(-i)v_n\overleftrightarrow{\partial_x}v^*_n+\frac{b^*\overleftrightarrow{\partial_x}b}{2i}\right),
\end{split}
\end{equation}
where $E_{\psi}$ is the eigenvalue of the mean field Hamiltonian $\hat{\mathcal{H}}_{\psi}$ in \req{eq:meanpsi} under this particular filling configuration, and the double arrow derivative is defined as
\begin{equation}
	f\overleftrightarrow{\partial_x}g\equiv f(\partial_xg)-(\partial_xf)g.
\end{equation}

The solutions to \req{eq:eqforuv} have a special particle-hole symmetry that if $(u_n,v_n)^\text{T}$ is a solution with eigenvalue $\epsilon_n$, then $(-v^*_n,u^*_n)^\text{T}$ must be a solution with eigenvalue $-\epsilon_n$. As a result, nonzero eigenvalues appear in pairs. Moreover, if \req{eq:eqforuv} possesses zero eigenvalue, it must be degenerate, otherwise we would have
\begin{equation}
\label{eq:fanzheng}
	\begin{pmatrix}
		u_0 \\ v_0
	\end{pmatrix}=c\begin{pmatrix}
		-v^*_0 \\ u^*_0
	\end{pmatrix},
\end{equation}
where $(u_0,v_0)^{\text{T}}$ is the solution to \req{eq:eqforuv} for $\epsilon=0$ and $c$ is constant complex number of modulus one $|c|=1$. Equation (\ref{eq:fanzheng}) would then lead to $|c|^2u_0=-u_0$, which cannot be true unless $u_0$ is trivially zero (This argument is analogous to that for Kramers degeneracy).  In later sections where $\Delta(x,t)$ is identified as a soliton, we find that the degenerate zero modes appear only in the deep BCS limit, where the spectrum is linearized around the Fermi points. But this turns out to be an artifact of the linearization, and there will be no zero mode when the nonlinear effect of the spectrum is taken into account.

It is clear from the above analysis that the solutions to \req{eq:eqforuv} always appear in pairs, the state $S=0$ then corresponds to a zero (or even) occupation of Bogoliubov fermions $\hat{\gamma}_{n\sigma}$ and the state $S=1/2$ is made out of odd occupation. Also as we will see in later sections, the state of the $S=0$ soliton corresponding to the exact solution is not necessarily a ground state of $\hat{\mathcal{H}}_{\psi}$.

\subsection{Dark Soliton}
The dark soliton is characterized by a twist in the configuration of $b(x)$ where its value changes sign rapidly from $x<0$ to $x>0$. Taking into consideration the periodic boundary condition, $b(x)$ then has the following asymptotic behavior at spatial boundaries:
\begin{equation}
	b(x\to \pm L/2) \sim e^{ i\pi x/L},
\end{equation}
where we are taking the infinite system limit that $L\to \infty$. It would be helpful to perform the following gauge transformation:
\begin{equation}
\label{eq:pigauge}
	b(x) = e^{ i \pi x/L}\tilde{b}(x),
\end{equation}
then the dark soliton can be presented as
\begin{equation}
	\tilde{b}(x)=-ib_0f\left(\frac{x}{l_s}\right),
\end{equation}
where $l_s\ll L$ is the size of the soliton sitting at $x=0$, the constant number $b_0$ is chosen to be real, and the shape function $f(x)$ has the asymptotic behavior that $f(x\to\pm \infty)=\pm 1$. Under this gauge transformation, the periodic boundary condition of $b(x)$ becomes $\tilde{b}(x+L)=-\tilde{b}(x)$. As a result, $\tilde{b}(x)$ can be chosen purely imaginary, or equivalently, $f(x)$ can be chosen purely real.

To get rid of the phase in \req{eq:pigauge}, we perform the following gauge transformation on the classical fields $u_n(x), v_n(x)$:
\begin{equation}
	\begin{cases}
		u_n(x)=e^{ i\pi x/L}\tilde{u}(x) \\
		v_n(x)=\tilde{v}_n
	\end{cases},
\end{equation}
then \req{eq:eqforuv} is transformed into
\begin{equation}
\label{eq:gaugeequv}
	\begin{split}
	& \begin{pmatrix}
		-\partial^2_x-\mu & t_b\tilde{b} \\ (t_b\tilde{b})^* & \partial^2_x+\mu
	  \end{pmatrix}\begin{pmatrix}
		\tilde{u}_n \\ \tilde{v}_n
	  \end{pmatrix}=\epsilon_n\begin{pmatrix}
		\tilde{u}_n \\ \tilde{v}_n
	  \end{pmatrix}, \\
	& ~~\begin{cases}\tilde{u}_n(x+L,t)=-\tilde{u}_n(x,t)\\ \tilde{v}_n(x+L,t)=\tilde{v}_n(x,t)\end{cases},
\end{split}
\end{equation}
where we have neglected both $L^{-1}$ and $L^{-2}$ correction to the eigenenergy $\epsilon_n$. The former can be neglected because it contributes to the total energy in \req{eq:exforEP1} a term proportional to $P/L$, which goes to zero in the limit $L\to \infty$ for finite momentum $P$. The latter can be neglected because it contributes to the total energy a term proportional to $NL^{-2}$, which also goes to zero in the limit $L\to \infty$.  Using these gauge transformed classical fields, the energy $E$, the momentum $P$ and the conserved quantity $N$ of the system can be expressed as:
\begin{equation}
\label{eq:exforEP2}
\begin{split}
	E=&\int dx\left( \frac{1}{2}|\partial_x\tilde{b}|^2-(2\mu+\epsilon_b)|\tilde{b}|^2 \right)+E_{\psi},\\
	P=& \int dx\left(\sum_{\epsilon_n>0}\frac{\tilde{u}^*_n\overleftrightarrow{\partial_x}\tilde{u}_n+\tilde{v}^*_n\overleftrightarrow{\partial_x}\tilde{v}_n}{2i}\sum_{\sigma}\lrangle{\hat{\gamma}^{\dagger}_{n\sigma}\hat{\gamma}_{n\sigma}} \right)\\
	&+\int dx\left(\sum_{\epsilon_n>0}(-i)\tilde{v}_n\overleftrightarrow{\partial_x}\tilde{v}^*_n+\frac{\tilde{b}^*\overleftrightarrow{\partial_x}\tilde{b}}{2i}\right)+\frac{N}{2L}\pi,\\
	N=&\int dx\sum_{\epsilon_n>0}\left[(\tilde{u}^*_n\tilde{u}_n-\tilde{v}^*_n\tilde{v}_n)\sum_{\sigma}\lrangle{\hat{\gamma}^{\dagger}_{n\sigma}\hat{\gamma}_{n\sigma}}\right]\\
	&+\int dx\left(
	\sum_{\epsilon_n>0}2\tilde{v}^*_n\tilde{v}_n+2\tilde{b}^*\tilde{b}\right),
\end{split}
\end{equation}
where the effect of gauge transformation is taken into account in the limit $L\to \infty$. The contribution to the energy is vanishingly small ($\sim NL^{-2}$) while the contribution to the momentum remains finite, which appears in the expression for $P$ as the last term proportional to $n=N/L$.

To be consistent with the choice that $\tilde{b}(x)$ is purely imaginary, $\tilde{u}_n(x)$ and $\tilde{v}_n(x)$ can be chosen purely real and purely imaginary respectively. Since the classical fields $\tilde{b}(x)$, $\tilde{u}_n(x)$ and $\tilde{v}_n(x)$ are chosen to be either purely real or purely imaginary, the contribution from the integral for the momentum $P$ in \req{eq:exforEP2} is zero, then we arrive at the result that the momentum of the dark soliton is exactly the Fermi momentum:
\begin{equation}
	P=k_F=\pi n/2,
\end{equation}
whether it is for a $S=0$ state or a $S=1/2$ state.

Now we have to determine the actual form of the dark soliton profile $f(x)$. It is obtained by solving the equation of motion for the classical field $\tilde{b}(x)$. Because the dark soliton corresponds to a local minimum or a local maximum of the energy for $S=0$ or $S=1/2$ spectrum respectively, the desired equation of motion for $\tilde{b}(x)$ can be derived by extremizing the energy $E$ in \req{eq:exforEP2}:
\begin{equation}
\label{eq:semi0}
	-\frac{1}{2}\partial^2_x \tilde{b}-\left( 2\mu+\epsilon_b \right)\tilde{b}+\frac{\delta E_{\psi}}{\delta \tilde{b}^*}=0.
\end{equation}
Together with \req{eq:gaugeequv}, we now have a complete set of equations to determine all the relevant classical fields.

As mentioned at the end of section II, our proposal for $S=1/2$ excitations is based upon the assumption that one extra fermion can be bounded on the dark soliton, which is equivalent to the assumption that there is at least one localized state solution to \req{eq:gaugeequv}, so we present below a simple one-parameter variational approach to verify this assumption.

The Hamiltonian operator corresponding to \req{eq:gaugeequv} is as follows:
\begin{equation}
	\hat{\mathcal{H}}_b=\begin{pmatrix}
		-\partial^2_x-\mu & t_b\tilde{b} \\ (t_b\tilde{b})^* & \partial^2_x+\mu
	  \end{pmatrix},
\end{equation}
and it has a positive as well as a negative sector, due to the particle-hole symmetry discussed after \req{eq:eqforuv}. Accordingly, the existence of the localized state can be proved by the fact that the expectation value $I(\kappa)$ of $\hat{\mathcal{H}}_b^2$ on a normalized trial wave function $\psi_{\kappa}(x)$ is below the boundary of the continuous spectrum for $\hat{\mathcal{H}}^2_b$, where $\kappa$ is the variational parameter:
\begin{equation}
\begin{split}
	& I(\kappa)=\int dx\left(\mathcal{H}\psi_{\kappa}(x) \right)^*\mathcal{H}\psi_{\kappa}(x), \\
	& \int dx ~\psi^*_{\kappa}(x)\psi_{\kappa}(x)=1.
\end{split}
\end{equation}
Here we make the choice that $I(0)$ corresponds to the boundary of the continuous spectrum and $\kappa>0$ corresponds to the localized state. Then the existence of localized state corresponds to $I'(0)<0$.

For $\mu>0$, the boundary of the continuous spectrum for $\hat{\mathcal{H}}_b^2$ is $\Delta_0^2=(t_bb_0)^2$, and the normalized trial wave function can be chosen as
\begin{equation}
	\psi_{\kappa}=\sqrt{\kappa}e^{-\kappa|x|}\begin{pmatrix}
		\cos k_Fx \\ \sin k_Fx
	\end{pmatrix},
\end{equation}
where $k_F^2=\mu$. Then we have
\begin{equation}
	\!\!I(\kappa)=\Delta^2_0+\kappa^4+4\kappa^2k^2_F-\Delta^2_0\kappa l_s\kern-0.5em\int e^{-2\kappa l_s|y|}f^2(y)dy,
\end{equation}
which has the required property that
\begin{equation}
	I(0)=\Delta^2_0, ~~~ I'(0)<0.
\end{equation}
For $\mu\leqslant 0$, the boundary of the continuous spectrum for $\hat{\mathcal{H}}^2_b$ is $\Delta^2_0+\mu^2$, and the following normalized trial wave function is chosen:
\begin{equation}
	\psi_{\kappa}=\sqrt{\kappa}e^{-\kappa|x|}\begin{pmatrix}
		1 \\ 0
	\end{pmatrix},
\end{equation}
Then we have
\begin{equation}
\kern-0.5emI(\kappa)=(-\kappa^2\!+|\mu|)^2+\Delta^2_0-\Delta^2_0\kappa l_s\kern-0.5em\int e^{-2\kappa l_s|y|}f^2(y)dy,
\end{equation}
which again has the required property that
\begin{equation}
	I(0)=\Delta^2_0+\mu^2,~~~ I'(0)<0.
\end{equation}
Taking also into consideration that the solutions to \req{eq:gaugeequv} always appear in pairs and belong to the negative and positive sectors respectively, we then proved here that there is at least one localized state for each sector for the whole range of $\mu$ across the BCS-BEC crossover.

In later sections, we will show that the number of localized state is exactly one for each sector in both the deep BCS and the deep BEC limit, and we didn't find any evidence for the existence of a second localized state (appearance of such state would not violate any further consideration).

\subsection{Grey Soliton}
In order to transform the dark soliton into a moving grey soliton, we need to generalize the above construction to the following asymptotic behavior at spatial boundaries:
\begin{equation}
\label{eq:boundarycon}
	b(x\to \pm L/2,t) \sim e^{ i\theta_s x/L},
\end{equation}
where the phase parameter $\theta_s\in [0,2\pi)$ and we take the limit $L\to \infty$. We will show in later sections that the moving grey soliton can be presented in the following form:
\begin{equation}
\label{eq:solsolu}
	b(x,t)=\left[ \cos\frac{\theta_s}{2}-i\sin\frac{\theta_s}{2}f\left(\frac{x-v_st}{l_s}\right)\right]e^{ i\theta_s x/L},
\end{equation}
where $v_s$ is the velocity of the grey soliton. The velocity $v_s$ and phase parameter $\theta_s$ are not independent variational variables. As we will show now, they are related to each other via the semiclassical velocity formula $v_s=\partial E(\theta_s)/\partial P(\theta_s)$.

Considering the transformation of the variables from $(x,t)$ to $(z,t)$ such that $z=x-v_st$, we will have
\begin{equation}
\label{eq:veltrans}
\begin{split}
& \hat{\mathcal{H}} \to \hat{\Omega}=\hat{\mathcal{H}}+\frac{iv_s}{2}\int dz\left(\hat{\psi}^{\dagger}\overleftrightarrow{\partial_z}\hat{\psi}+\hat{b}^{\dagger}\overleftrightarrow{\partial_z}\hat{b} \right),\\
& \hat{\mathcal{P}} \to \hat{\mathcal{P}}=\frac{(-i)}{2}\int dz\left(\hat{\psi}^{\dagger}\overleftrightarrow{\partial_z}\hat{\psi}+\hat{b}^{\dagger}\overleftrightarrow{\partial_z}\hat{b} \right),
\end{split}
\end{equation}
where we have variable $x$ on the lefthand side and variable $z$ on the righthand side. We can see that in \req{eq:veltrans} new terms are added to the Hamiltonian operator, while the momentum operator remains unchanged. This implies that the variable transformation introduced here is not a Galilean transformation, for which the momentum would have been changed by the amount proportional to $v_sN\to \infty$. From \req{eq:veltrans} we obtain the following operator relations:
\begin{equation}
	\frac{\partial \hat{\Omega}}{\partial v_s}=-\hat{\mathcal{P}}, ~~~ \hat{\Omega}=\hat{\mathcal{H}}-v_s\hat{\mathcal{P}}.
\end{equation}
By taking the expectation values of both sides on the soliton with phase parameter $\theta_s$, we can see that the change of variables from $x$ to $z$ is equivalent to a Legendre transformation:
\begin{equation}
\label{eq:legleg}
	\frac{\partial \Omega(\theta_s)}{\partial v_s}=-P(\theta_s), ~~~ \Omega(\theta_s)=E(\theta_s)-v_sP(\theta_s).
\end{equation}
By taking derivative with respect to $\theta_s$ of both sides of the second equation in (\ref{eq:legleg}), we obtain
\begin{equation}
\label{eq:leg1}
	\frac{\partial \Omega}{\partial \theta_s}=\frac{\partial E}{\partial\theta_s}-\frac{\partial v_s}{\partial \theta_s}P-v_s\frac{\partial P}{\partial\theta_s}.
\end{equation}
Then using the first equation in (\ref{eq:legleg}) we also obtain
\begin{equation}
\label{eq:leg2}
	\frac{\partial \Omega}{\partial \theta_s}=\frac{\partial \Omega}{\partial v_s}\frac{\partial v_s}{\partial \theta_s}=-\frac{\partial v_s}{\partial \theta_s}P.
\end{equation}
Combining \req{eq:leg1} and \req{eq:leg2}, we arrive at the following equation
\begin{equation}
\label{eq:vel}
	v_s=\frac{\partial E/\partial \theta_s}{\partial P/\partial \theta_s}=\frac{\partial E}{\partial P}.
\end{equation}
When the soliton is interpreted as a proper excitation, \req{eq:vel} is just the semiclassical velocity formula mentioned above, which determines the soliton velocity $v_s$ as a function of $\theta_s$. Then the fact that the dark soliton corresponds to either the maximum ($S=0$) or minimum energy ($S=1/2$) follows from the condition that $v_s(\theta_s=\pi)=0$.

As in derivation for the dark soliton, it would be helpful to do the following gauge transformation of the classical fields:
\begin{equation}
\label{eq:gaugetrans}
\begin{split}
	& b(x,t)=e^{ i\frac{\theta_s x}{L}}\tilde{b}(z),\\
	& u_n(x,t)=e^{ i\frac{\theta_s x}{L}}\tilde{u}_n(z), \\
	& v_n(x,t)=\tilde{v}_n(z),
\end{split}
\end{equation}
where $z=x-v_st$. This leaves us with the analysis of classical fields $\tilde{b}(z)$ or $\tilde{\Delta}(z)=t_b\tilde{b}$(z), $\tilde{u}_n(z)$ and $\tilde{v}_n(z)$, for which we will omit the tilde in the following whenever there is no confusion. Also, the gauge transformation modifies the boundary conditions of the classical fields:
\begin{equation}
\label{eq:boundCon}
	\begin{split}
		& b(z+L)=e^{-i\theta_s}b(z), \\
		& u_n(z+L)=e^{-i\theta_s}u_n(z), \\
		& v_n(z+L)=v_n(z).
	\end{split}
\end{equation}
Using these classical fields, again we can write down the expressions for the energy $E$, momentum $P$ and the conserved quantity $N$:
\begin{equation}
\label{eq:EPN}
\begin{split}
E=&\int dz\left( \frac{1}{2}|\partial_zb|^2-(2\mu+\epsilon_b)|b|^2 \right)+E_{\psi},\\
P=& \int dz\left(\sum_{\epsilon_n>0}\frac{u^*_n\overleftrightarrow{\partial_z}u_n+v^*_n\overleftrightarrow{\partial_z}v_n}{2i}\sum_{\sigma}\lrangle{\hat{\gamma}^{\dagger}_{n\sigma}\hat{\gamma}_{n\sigma}} \right)\\
	&+\int dz\left(\sum_{\epsilon_n>0}(-i)v_n\overleftrightarrow{\partial_z}v^*_n+\frac{b^*\overleftrightarrow{\partial_z}b}{2i}\right)+\frac{N}{2L}\theta_s,\\
N=&\int dz\sum_{\epsilon_n>0}\left[(u^*_nu_n-v^*_nv_n)\sum_{\sigma}\lrangle{\hat{\gamma}^{\dagger}_{n\sigma}\hat{\gamma}_{n\sigma}}\right]\\
	&+\int dz\left(
	\sum_{\epsilon_n>0}2v^*_nv_n+2b^*b\right),
\end{split}
\end{equation}
where the energy and momentum are understood by taking the reference point that $E(\theta_s=0)=0$ and $P(\theta_s=0)=0$. Also, the chemical potential is determined by the usual thermodynamic relation that $\mu=\partial E/\partial N$.

\begin{figure}[htp!]
\includegraphics[scale=0.3]{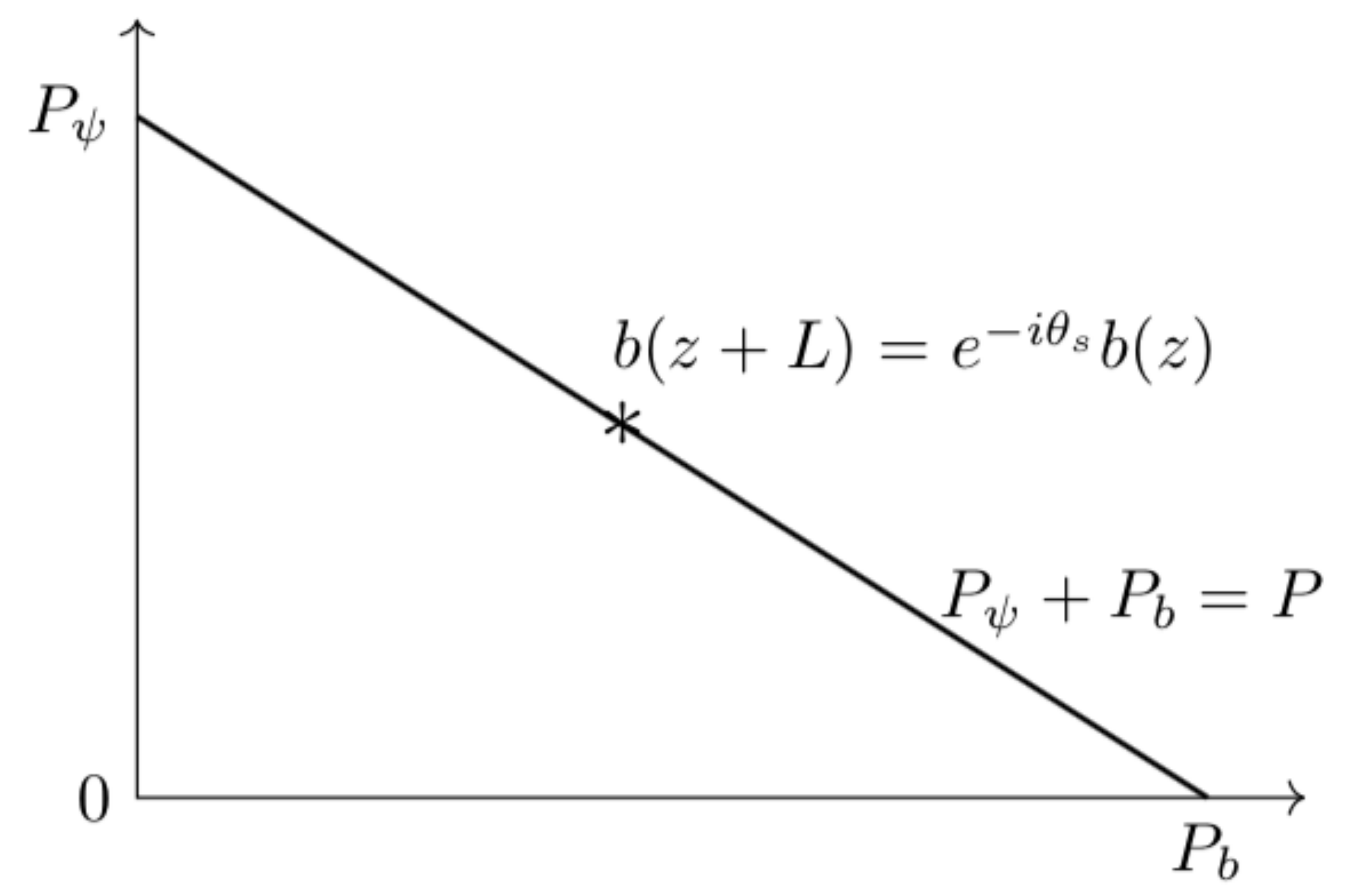}
\caption{\footnotesize The functional space for the extremization spanned by $P_{\psi}$ and $P_{b}$. The thick line is the collection of extreme points and the starred point is the one that satisfy the required boundary condition in \req{eq:boundCon}.}
\label{fig:extremize}
\end{figure}

For a particular filling configuration of the mean field Hamiltonian $\hat{\mathcal{H}}_{\psi}$ in \req{eq:meanpsi}, we now derive the semiclassical equations of motion for the classical fields $b(z), u_n(z)$ and $v_n(z)$. Unlike the dark soliton, the grey soliton only extremizes the energy $E$ under certain constraints. Usually we would extremize the energy $E$ under the constraint of fixed momentum $P$, but this approach may not respect the desired boundary condition in \req{eq:boundCon}. To overcome this difficulty, we use a modified extremization process. Firstly, we partition the momentum $P$ in \req{eq:EPN} into two parts: the contribution $P_{\psi}$ from the fermion fields and the contribution $P_b$ from the $b$ field:
\begin{equation}
		P_b=\int dz\left(\frac{(-i)}{2}b^*\overleftrightarrow{\partial_z}b+\frac{b^*b}{L}\theta_s\right), ~~~ P_{\psi}=P-P_b
\end{equation}
 Then instead of keeping $P$ fixed, we keep both $P_{\psi}$ and $P_{b}$ fixed, and this introduces two Lagrangian multiplier $v_{\psi}$ and $v_b$ into the free energy $F$ we want to extremize:
\begin{equation}
\label{eq:freeenergyF}
	E \to F=E-v_{\psi}P_{\psi}-v_bP_b.
\end{equation}
We can visualize this modified extremization in the functional space spanned by $P_{\psi}$ and $P_{b}$ (see Fig. \ref{fig:extremize}). Each point on the hyperline $P_{\psi}+P_b=P$ corresponds to an extreme of the free energy $F$, and one point among them (the starred point in Fig. \ref{fig:extremize}) is picked out by adjusting the Lagrangian multiplier pair $(v_{\psi},v_b)$ to satisfy the boundary condition in \req{eq:boundCon}. This modified extremization process is morally equivalent to the method of constrained instanton used in field theories \cite{AFFLECK1981429}. Also, following the derivation from \req{eq:legleg} to \req{eq:vel}, we obtain
\begin{equation}
\label{eq:vel1}
	dE=v_{\psi}dP_{\psi}+v_bdP_b=v_s(dP_{\psi}+dP_b).
\end{equation}
This allows a trivial solution that $v_{\psi}=v_b=v_s$ or a nontrivial solution such that
\begin{equation}
\label{eq:vel2}
	 \frac{v_s-v_{\psi}}{v_b-v_s}=\frac{\partial P_b}{\partial P_{\psi}}.
\end{equation}
We will see in later sections that the nontrivial solution is crucial on the deep BCS side.

Applying the modified extremization process, we obtain the following equations of motion for the classical fields in the limit $L\to \infty$:
\begin{equation}
\label{eq:semi1}
-\frac{1}{2}\partial^2_z b+iv_b\partial_zb-\left( 2\mu+\epsilon_b \right)b+\frac{\delta E_{\psi}}{\delta b^*}=0,
\end{equation}
\begin{equation}
\label{eq:semi2}
	\begin{pmatrix}-\partial^2_z-\mu+iv_{\psi}\partial_z & \Delta(z) \\ \Delta^*(z) & \partial^2_z+\mu+iv_{\psi}\partial_z\end{pmatrix}\begin{pmatrix}u_n \\ v_n\end{pmatrix}=\bar{\epsilon}_n \begin{pmatrix}u_n \\ v_n\end{pmatrix},
\end{equation}
where $\Delta(z)$ field is related to $b(z)$ field through the definition $\Delta(z)=t_bb(z)$ and the eigenvalue $\bar{\epsilon}_n$ differs from $\epsilon_n$ in \req{eq:meanpsi} and \req{eq:eqforuv} in that $\bar{\epsilon}_n$ contributes to the free energy $F$ in \req{eq:freeenergyF} while $\epsilon_n$ contributes to the energy $E$ in \req{eq:exforEP1}. We should keep this in mind when later calculating the energy $E$. Also, the proof of the existence of the localized state for a dark soliton can be easily generalized here to \req{eq:semi2} for a grey soliton.

\section{Theory of $S=1/2$ Soliton}
In this section, we apply the general formalism outlined above to the $S=1/2$ soliton, which turns out to be simpler than the $S=0$ soliton. The two weak coupling limits - the deep BCS side and the deep BEC side - permit analytical treatment, because on either side, one of the degrees of freedom lies high in energy compared to the other such that we are left with a decoupled theory with weak interaction.

\subsection{Deep BCS Side}
On the deep BCS side, we tune the resonant level of $b$ field far above the Fermi sea such that $\epsilon_b<0, |\epsilon_b|\gg \mu$. Since the $b$ field now only acts as a virtual state to effect the low energy physics, we can ignore its dynamics, and the equation of motion for it reduces to a self-consistent equation:
\begin{equation}
\label{eq:selfcon}
	\Delta=\lambda\sum_{\epsilon_n>0}u_{n}v_{n}^*\left(1-\sum_{\sigma}\lrangle{\hat{\gamma}_{n\sigma}^{\dagger}\hat{\gamma}_{n\sigma}}\right)+\tau(i\partial_z\Delta),
\end{equation}
where $\lambda=\frac{|t_b|^2}{-(2\mu+\epsilon_b)}>0$ serves as the effective coupling constant and $\tau=v_b/(2\mu+\epsilon_b)$. Also for the dark soliton, we should bear in mind that we need to set $v_{\psi}=v_b=0$ and $\tau=0$. Combined with the equation of motion for the fermion fields, we can reconstruct the Hamiltonian as
\begin{equation}
\label{eq:BCS1}
\begin{split}
\hat{\mathcal{H}}=&\int dz\left( \sum_{\sigma}\hat{\psi}^{\dagger}_{\sigma}\left( -\partial^2_z-\mu \right)\hat{\psi}_{\sigma}\right)\\
&+\int dz\left(\Delta^*\hat{\psi}_{\downarrow}\hat{\psi}_{\uparrow}+\Delta\hat{\psi}^{\dagger}_{\uparrow}\hat{\psi}^{\dagger}_{\downarrow}+\frac{|\Delta|^2}{\lambda}\right).
\end{split}
\end{equation}
This is just the BCS mean field Hamiltonian for the conventional superconductivity and the $\Delta$ field is just the gap parameter. The system is made up with loosely bounded Cooper pairs, and we have a large chemical potential $\mu=k^2_F$, where $k_F=\pi n/2$ and $n=N/L$. Since the low energy physics happens only near the two Fermi points, we can linearize the spectrum around them:
\begin{equation}
\label{eq:LRD}
\begin{pmatrix}u_n\\ v_n\end{pmatrix}=\sum_{\alpha}\begin{pmatrix}u^{\alpha}_n\\ v^{\alpha}_n\end{pmatrix}e^{i\alpha k_Fz},
\end{equation}
where $\alpha=-1$ and $\alpha=1$ denotes the left and right moving modes respectively. Correspondingly, \req{eq:semi2} can be linearized to the following form:
\begin{equation}
\label{eq:LinearEq}
\!\!\begin{pmatrix}-i\alpha v_F\partial_z-\alpha v_{\psi}k_F & \kern-1em \Delta(z)\\ \Delta^*(z) & \kern-1em i\alpha v_F\partial_z-\alpha v_{\psi}k_F \end{pmatrix}\begin{pmatrix}u^{\alpha}_n\\ v^{\alpha}_n\end{pmatrix}=\bar{\epsilon}^{\alpha}_n\begin{pmatrix}u^{\alpha}_n\\ v^{\alpha}_n\end{pmatrix},
\end{equation}
where the bar notation of the eigenvalue again reminds us that $\bar{\epsilon}^{\alpha}_n$ contributes to the free energy $F$ instead of the energy $E$. Moreover, due to the linearization made here, we can further determine the eigenvalue $\epsilon^{\alpha}_n$ that contributes to energy $E$ as $\epsilon^{\alpha}_n=\bar{\epsilon}^{\alpha}_n+\alpha v_{\psi}k_F$.

The solution to this linearized Bogoliubov-de Gennes equation under soliton profile in the context of polyacetylene and charge density waves is well established in the literature \cite{PhysRevB.22.2099,PhysRevB.21.2388,Brazovskii_1989,1980JETP...51..342B}. Essentially, the solvability comes from the fact that \req{eq:LinearEq} has the form of Dirac equation in one dimension and it can be associated with a nonlinear Schr$\ddot{\text{o}}$dinger equation for the $\Delta(z)$ field via the inverse scattering method \cite{Faddeev_ISM}.  There then exists the following soliton solution:
\begin{equation}
\label{eq:FSoliton}
\Delta(z)=\Delta_0\left[ \cos\frac{\theta_s}{2}-i\sin\frac{\theta_s}{2}\tanh\left( \frac{z}{l_s} \right) \right],
\end{equation}
where the size of the soliton is $l^{-1}_s=\left(\Delta_0\sin\frac{\theta_s}{2}\right)/v_F$. The eigenmodes of \req{eq:LinearEq} can be classified into two categories. The first category includes the delocalized states labelled by left-right moving index $\alpha=\pm$, band index $\iota=\pm$ and momentum $k$:
\begin{equation}
\label{eq:deEigen}
\begin{split}
& \begin{cases}
u^{\alpha}_{\iota k}=\frac{1}{2}\frac{1}{\sqrt{N^{\alpha}_{\iota k}L}}\left[ 1+\alpha\frac{v_Fk+i\Delta_2\tanh\left(\frac{\Delta_2}{v_F}z \right)}{\epsilon_{\iota k}-\alpha \Delta_1} \right]e^{ikz} \\
v^{\alpha}_{\iota k}=\frac{1}{2}\frac{1}{\sqrt{N^{\alpha}_{\iota k}L}}\left[ -\alpha+\frac{v_Fk+i\Delta_2\tanh\left(\frac{\Delta_2}{v_F}z \right)}{\epsilon_{\iota k}-\alpha \Delta_1} \right]e^{ikz}
\end{cases}, \\
& \Delta_1=\Delta_0\cos\frac{\theta_s}{2}, \Delta_2=\Delta_0\sin\frac{\theta_s}{2}, N^{\alpha}_{\iota k}=\frac{\epsilon_{\iota k}}{\epsilon_{\iota k}-\alpha \Delta_1}.
\end{split}
\end{equation}
The corresponding eigenvalues are
\begin{equation}
\label{eq:deEigenE}
	\bar{\epsilon}^{\alpha}_{\iota k}=\epsilon_{\iota k}-\alpha v_{\psi}k_F, ~ \epsilon_{\iota k}=\iota\epsilon_k,~\epsilon_k=\sqrt{\Delta^2_0+v^2_Fk^2}.
\end{equation}
so the band $\iota=+$ corresponds to the excitations defined in \req{eq:meanpsi}. The second category is the localized states on the soliton core, labelled only by the left-right moving index $\alpha$:
\begin{equation}
\label{eq:local1}
	\begin{pmatrix}u^{\alpha}_0\\ v^{\alpha}_0\end{pmatrix}=\frac{1}{2}\sqrt{\frac{\Delta_2}{v_F}}\text{sech}\left( \frac{\Delta_2z}{v_F} \right)\begin{pmatrix}1 \\ \alpha\end{pmatrix},
\end{equation}
and the corresponding eigenvalues are:
\begin{equation}
\label{eq:local2}
	\bar{\epsilon}^{\alpha}_0=\epsilon^{\alpha}_0-\alpha v_{\psi}k_F, ~~~ \epsilon^{\alpha}_0=\alpha\Delta_0\cos\frac{\theta_s}{2}.
\end{equation}
According to the above expression for the eigenvalues, the localized states corresponding to the dark soliton ($\theta_s=\pi, v_{\psi}=0$) are degenerate zero modes, but this degeneracy is an artifact of the linearization in \req{eq:LRD}, while the correction from the quadratic spectrum up to leading order will lift this degeneracy:
\begin{equation}
\label{eq:localeigen}
	\epsilon^{\alpha}_0=\alpha \sqrt{\Delta^2_0-\left[\Delta_0\sin\frac{\theta_s}{2}-\frac{\pi v_Fk_F}{2}\text{csch}\left(\frac{\pi l_sk_F}{2} \right)\right]^2}.
\end{equation}
The actual localized states are linear combinations of the left and right moving localized states, so the superscript $\alpha=\pm$ in \req{eq:localeigen} labels positive and negative modes instead of left and right moving modes.

To complete the construction of the soliton, we still need to satisfy the self-consistent requirement in \req{eq:selfcon}. In the present classification of the eigenmodes, it is expressed as
\begin{equation}
\label{eq:selfcon22}
\begin{split}
\Delta=\lambda\sum_{\alpha,k}u^{\alpha}_{+,k}{v^{\alpha}_{+,k}}^*\left(1-\sum_{\sigma}\lrangle{\hat{\gamma}^{\alpha\dagger}_{+,k,\sigma}\hat{\gamma}^{\alpha}_{+,k,\sigma}}\right)\\+\lambda u^{+}_0{v^{+}_0}^*\left(1-\sum_{\sigma}\lrangle{\hat{\gamma}^{+\dagger}_{0,\sigma}\hat{\gamma}^{+}_{0,\sigma}} \right)+\tau (i\partial_z\Delta).
\end{split}
\end{equation}
The $S=1/2$ soliton is obtained by setting $\sum_{\sigma}\lrangle{\hat{\gamma}^{\alpha\dagger}_{+,k,\sigma}\hat{\gamma}^{\alpha}_{+,k,\sigma}}=0$ and $\sum_{\sigma}\lrangle{\hat{\gamma}^{+\dagger}_{0,\sigma}\hat{\gamma}^{+}_{0,\sigma}}=1$, then the above equation reduces to
\begin{equation}
\label{eq:selfcon2}
	\Delta=\lambda \int\frac{dk}{2\pi}\frac{\Delta}{\epsilon_k}+\frac{\lambda}{4}\frac{\Delta_0}{v_F}\frac{\theta_s-\pi}{\pi}\frac{\sin\frac{\theta_s}{2}}{\cosh^2\left(\frac{\Delta_2}{v_F}z \right)}+\tau(i\partial_z\Delta).
\end{equation}
For the dark soliton, the second part on the righthand side vanishes and we need to set $\tau=0$, then the resulting equation is exactly the one we have in conventional BCS theory with a homogenous gap parameter:
\begin{equation}
\label{eq:normalself}
	1=\lambda \int\frac{dk}{2\pi}\frac{1}{\epsilon_k} \Rightarrow \Delta_0\propto \exp\left(-\frac{1}{\lambda\nu(\epsilon_F)} \right),
\end{equation}
where $\nu(\epsilon_F)$ is the density of states on the Fermi level.

For the moving grey soliton, the second part on the righthand side of \req{eq:selfcon2} has a finite value, but it can be canceled by the third term under the choice that
\begin{equation}
\label{eq:tautau}
	\tau=\frac{\lambda}{4\Delta_0}\frac{\pi-\theta_s}{\pi}\sin^{-1}\frac{\theta_s}{2},
\end{equation}
which then determines $v_b$ as
\begin{equation}
\label{eq:vbbb}
	v_b=\frac{|t_b|^2}{4\Delta_0}\frac{\theta_s-\pi}{\pi}\sin^{-1}\frac{\theta_s}{2}.
\end{equation}
We can see that determination of parameters $\tau,v_b$ in the above equation is consistent with $\tau=0,v_b=0$ for $\theta_s=\pi$ in the case of dark soliton.

Having specified the $S=1/2$ soliton, we can proceed to calculate its energy and momentum near the dark soliton up to leading order in $\xi=\theta_s-\pi$ using the formula in \req{eq:EPN} and \req{eq:meanpsi}. The calculation consists of first determining the phase shift $\delta(k)$ for the continuous spectrum from the boundary conditions in \req{eq:boundCon} and then changing the summations over $k$ into integrations while taking into account the correction due to the phase shift $\delta(k)$ in the limit $L\to \infty$ \cite{PhysRevB.21.2388,PhysRevA.91.023616}. Also, we need to keep in mind that we should use $\epsilon^{\alpha}_n$ instead of $\bar{\epsilon}^{\alpha}_n$ in the calculation of energy $E$. The final result is:
\begin{equation}
\label{eq:Dispersion1}
\begin{split}
& E^{\text{BCS}}_{1/2}(\theta_s)=\frac{2\Delta_0}{\pi}\left(1+\frac{1}{8}\xi^2\right),\\
& P^{\text{BCS}}_{1/2}(\theta_s)=k_F-\frac{\Delta_0}{2v_F}\xi.
\end{split}
\end{equation}

\begin{figure}
\includegraphics[scale=0.5]{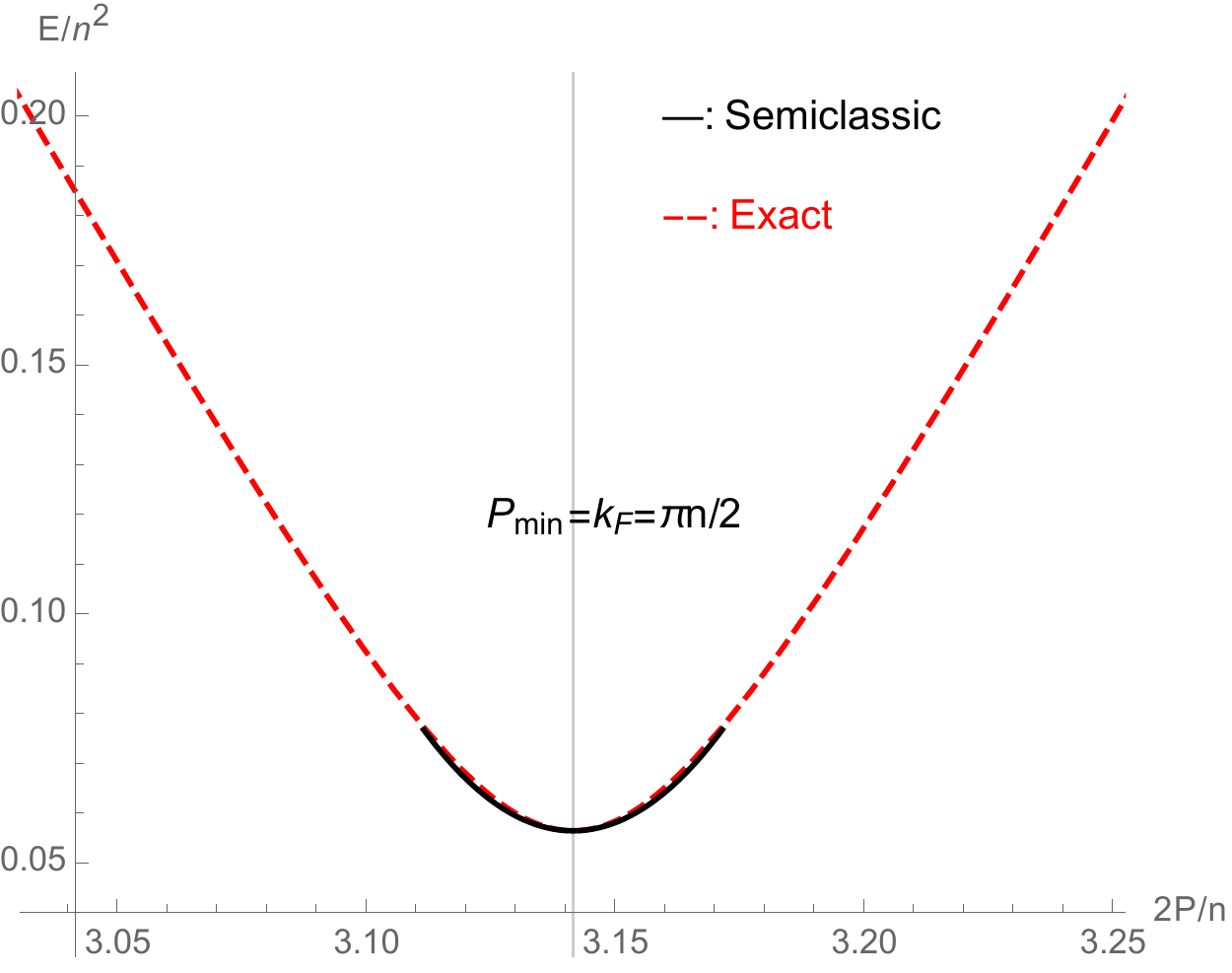}
\caption{\footnotesize The typical $S=1/2$ excitation spectrum in the semiclassical result and exact solution. The latter is plotted for $\gamma=c_F/n=1.13$, and correspondingly the former is plotted taking the spin gap at the same coupling strength as the input parameter.}
\label{fig:fermion2}
\end{figure}

This translates into the following dispersion relation and soliton velocity up to leading order in $\xi$:
\begin{equation}
\label{eq:velocity}
\begin{split}
& E_{1/2}=\frac{2\Delta_0}{\pi}\left(1+\frac{v^2_F(P_{1/2}-k_F)^2}{2\Delta^2_0} \right),\\
& v^\text{BCS}_s=\frac{\partial E_{1/2}}{\partial P_{1/2}}=-\frac{\xi}{\pi}v_F.
\end{split}
\end{equation}
It is clear that the minimum energy is achieved exactly at the Fermi momentum $k_F=\pi n/2$, as observed in the exact solutions. Also, the soliton velocity now is characterized by the Fermi velocity $v_F$, which is also consistent with the exact solutions. A comparison of the current semiclassical result with the exact solution is shown in Fig. \ref{fig:fermion2}, where the agreement is good in the vicinity of the dark soliton.

To complete the analysis, we still need to determine $v_{\psi}$ and $v_b$ from \req{eq:vel2}. In order to do that, we need the expressions for $P_{\psi}$ and $P_b$ respectively:
\begin{equation}
	P_b=\frac{\Delta^2_0}{|t_b|^2}(\pi+2\xi), ~~~ P_{\psi}=P^{\text{BCS}}_{1/2}(\theta_s)-P_b.
\end{equation}
Substituting them into \req{eq:vel2} and using \req{eq:vbbb}, we obtain up to leading order:
\begin{equation}
\label{eq:vbbb2}
	v_b=\frac{|t_b|^2}{4\pi\Delta_0}\xi, ~~~ v_{\psi}=v^{\text{BCS}}_s-\frac{v^{\text{BCS}}_s-v_b}{1+|t_b|^2/(4\Delta_0v_F)}\approx v^{\text{BCS}}_s,
\end{equation}
where the expression for $v_{\psi}$ will be of use in later section when we analyze the $S=0$ soliton. This closes our analysis of the $S=1/2$ soliton on the deep BCS side.

\subsection{Deep BEC Side}
On the BEC side, we tune the resonant level to a tightly bounded molecule with binding energy $\epsilon_b>0$. Then we have a negative chemical potential $\mu<0$ characterizing the absence of a Fermi sea, and we need to consider the quadratic Bogoliubov-de Gennes equation in \req{eq:semi2}. For delocalized states characterized by momentum $k$, we formally obtain the spectrum of Bogoliubov quasiparticles as:
\begin{equation}
\epsilon_k=\sqrt{(k^2-\mu)^2+|\Delta|^2}.
\end{equation}
For large negative chemical potential $\mu$, we can expanded the spectrum as
\begin{equation}
\epsilon_k=(k^2-\mu)+\frac{|\Delta|^2}{2(k^2-\mu)}-\frac{|\Delta|^4}{8(k^2-\mu)^3}+\cdots.
\end{equation}
Substituting this into \req{eq:semi1}, we can bring the equation of motion for $b=\Delta/t_b$ to the following form known as the Gross-Pitaevskii equation:
\begin{equation}
\label{eq:GP}
-\frac{1}{2}\partial^2_zb+iv_b\partial_zb+2g(|b|^2-n_s)b=0,
\end{equation}
where the parameters are defined via
\begin{equation}
g=\frac{3|t_b|^4}{128|\mu|^{5/2}}, ~~~ n_s=\frac{\frac{|t_b|^2}{4|\mu|^{1/2}}+2\mu+\epsilon_b}{\frac{3|t_b|^4}{64|\mu|^{5/2}}}.
\end{equation}

The Gross-Pitaevskii equation as a nonlinear Schr$\ddot{\text{o}}$dinger equation has been extensively studied in the literature \cite{Tsuzuki1971,1972JETP...34...62Z,1973JETP...37..823Z,Kulish:1976ek,PhysRevA.78.053630}. It also supports a soliton solution:
\begin{equation}
\label{eq:Soliton}
b(z)=\sqrt{n_s}\left( \cos\frac{\theta_s}{2}-i\sin\frac{\theta_s}{2}\tanh\frac{z}{l_s} \right),
\end{equation}
where  the size of the soliton is $l_s=[v_c\sin(\theta_s/2)]^{-1}$, the Lagrangian multiplier is $v_b=v_c\cos(\theta_s/2)$ and $v_c=\sqrt{gn}$ is the sound velocity. By calculating the total mass of the system we can also determine $n_s=n/2$.

\begin{figure}
\includegraphics[scale=0.4]{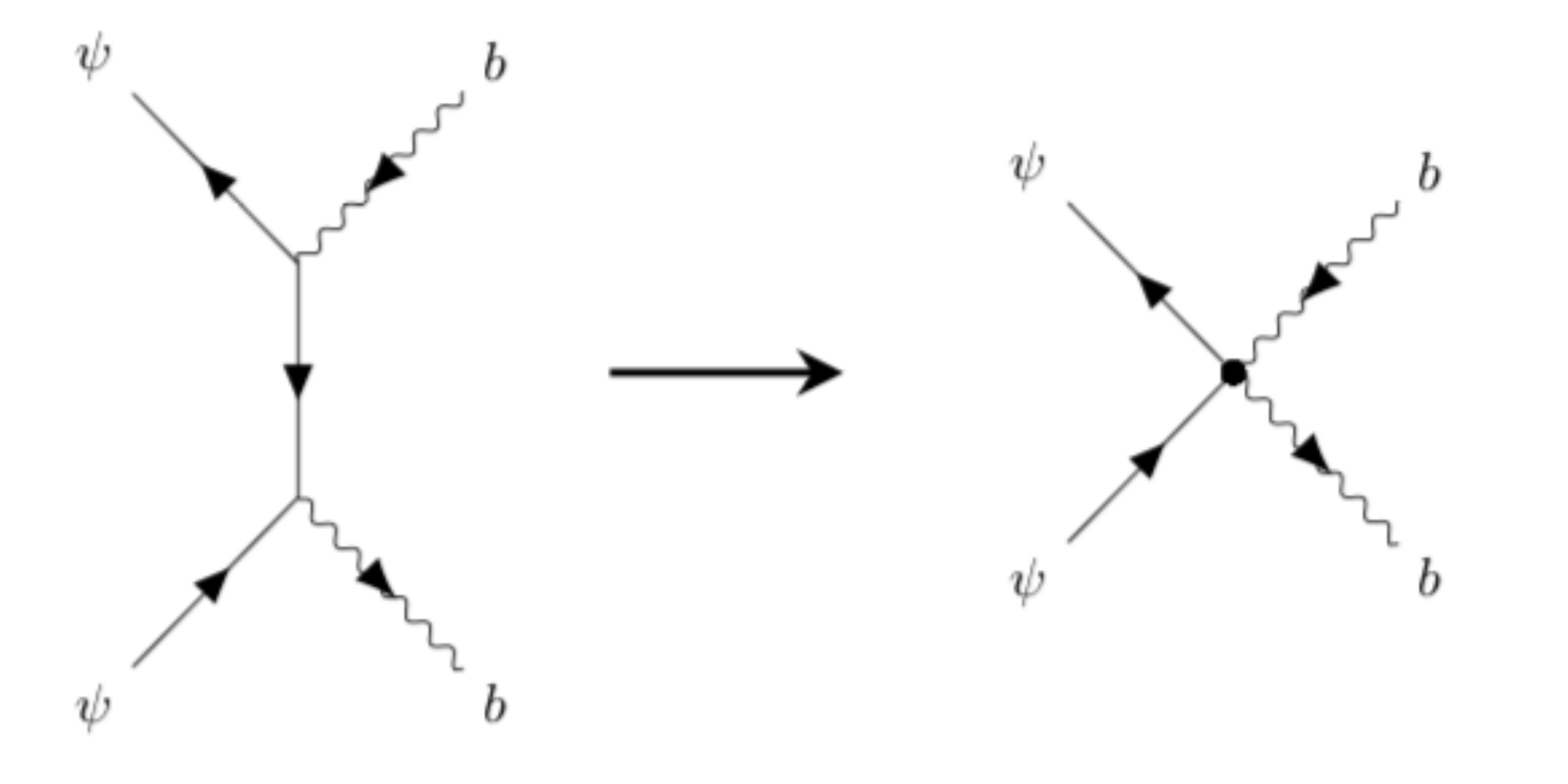}
\caption{\footnotesize The Feynman diagrams (left) for leading contribution to the scattering process $\psi b\to \psi b$ (right), where the solid line denotes the fermion propagator, the wiggled line denotes the boson propagator, the fermion-boson vertex denotes the resonant coupling $t_b$, and the dotted vertex on the right denotes the effective coupling $g'$.}
\label{fig:Feynman}
\end{figure}

The $S=1/2$ soliton is constructed by adding an extra fermion into the system, then we can effectively describe the system as follows: There is a weakly interacting  background (the bound pairs) with the effective coupling constant $g$. The extra fermion added into the system interacts with the background locally by an effective coupling constant $g'$, which can be calculated perturbatively from \req{eq:BCS-BEC} in the narrow resonance limit. For this purpose, we consider the scattering process $\psi b\to \psi b$, whose Feynman diagrams are shown in Fig. \ref{fig:Feynman}.
The scattering amplitude up to leading order is then
\begin{equation}
g'(\omega,k)=-\frac{|t_b|^2}{\omega-\epsilon_k}\approx \frac{|t_b|^2}{-2\mu}>0.
\end{equation}
As a result, the added fermion $\psi$ can be described as a quantum particle moving in the potential created by the background:
\begin{equation}
\label{eq:GP2}
(-\partial^2_z-\mu+iv_{\psi}\partial_z)\psi+g'(|b|^2-n_s)\psi=\bar{\epsilon}\psi,
\end{equation}
where in the second term on the lefthand side, we have adjusted for the interaction of the fermion with the uniform background (the constant term $g'n_s$), since it can be incorporated into the chemical potential. Performing the gauge transformation $\psi\to\psi e^{iv_{\psi}z/2}$ which shifts the momentum by $v_{\psi}/2$, and substituting \req{eq:Soliton} into \req{eq:GP2}, we end up with a Schr$\ddot{\text{o}}$dinger equation for a particle moving in the P$\ddot{\text{o}}$schl-Teller potential \cite{Poschl_Teller}:
\begin{equation}
-\partial^2_z\psi-\alpha^2\frac{\zeta(\zeta-1)}{\cosh^2\alpha z}\psi=\left(\bar{\epsilon}+\mu+\frac{v_{\psi}^2}{4}\right)\psi,
\end{equation}
where the two parameter $\alpha$ and $\zeta>1$ are determined by
\begin{equation}
\alpha=v_c\sin(\theta_s/2), ~~~ \alpha^2\zeta(\zeta-1)=g'n_s\sin^2(\theta_s/2).
\end{equation}
The P$\ddot{\text{o}}$schl-Teller potential produces a bound state with the following energy:
\begin{equation}
\begin{split}
\bar{\epsilon}_0& =-\alpha^2(\zeta-1)^2-\frac{v_{\psi}^2}{4}-\mu\\
&=-v_c^2\sin^2\frac{\theta_2}{2}\left( \frac{\sqrt{1+2g'/g}-1}{2} \right)^2-\frac{v_{\psi}^2}{4}+|\mu|.
\end{split}
\end{equation}
Also the momentum of this bound state is simply $k_0=v_{\psi}/2$, and we can determine the eigenvalue $\epsilon_0$ that contributes to the energy $E$ as
\begin{equation}
	\epsilon_0=\bar{\epsilon}_0+v_{\psi}k_0.
\end{equation}
Then the total energy $E_{1/2}(\theta_s)$ and momentum $P_{1/2}(\theta_s)$ of the system can be determined according to \req{eq:EPN}:
\begin{equation}
\label{eq:S2}
\begin{split}
E^{\text{BEC}}_{1/2}(\theta_s) &=\int dz~\left(\frac{1}{2}|\partial_zb|^2+g(|b|^2-n_s)^2\right)+\epsilon_0\\
& =n_sv_c\left[\frac{4}{3}\sin^3\frac{\theta_s}{2}-2u\sin^2\frac{\theta_s}{2} \right]+\frac{1}{4}v^2_{\psi}+|\mu|,\\
P^{\text{BEC}}_{1/2}(\theta_s) & =\int dz~\frac{1}{2i}(b^*\partial_zb-b\partial_zb^*)+n_s\theta_s+k_0\\
& = n_s(\theta_s-\sin\theta_s)+\frac{1}{2}v_{\psi},
\end{split}
\end{equation}
where $u=\sqrt{\frac{g}{n}}\left( \frac{\sqrt{1+2g'/g}-1}{2} \right)^2\gg 1$ in the narrow resonance limit. Thus the minimum of $E^{\text{BEC}}_{1/2}(\theta_s)$ is at $\theta_s=\pi$ with the following minimum energy:
\begin{equation}
E^\text{BEC}_{1/2}(\theta_s=\pi)-|\mu|=n_sv_c\left(\frac{4}{3}-2u\right)<0,
\end{equation}
so $E^{\text{BEC}}_{1/2}(\theta_s=\pi)$ is lower than the energy of adding one particle with zero momentum to the uniform background of bound pairs. Again, we arrive at the conclusion that the minimum energy is achieved exactly at the Fermi momentum $k_F=\pi n_s=\pi n/2$.

\begin{figure}
\includegraphics[scale=0.5]{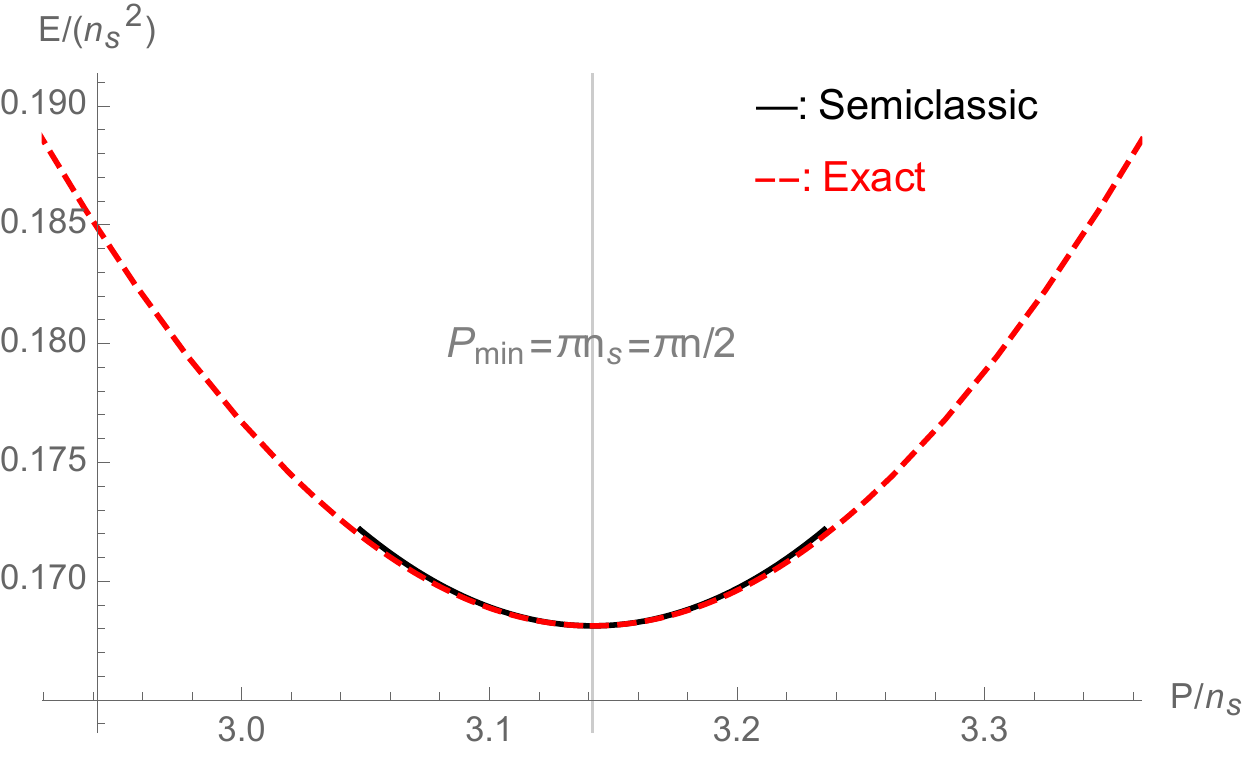}
\caption{\footnotesize The typical $S=1/2$ excitation spectrum in the semiclassical result and the exact solution. The former is plotted for $\gamma=g/n=0.07$, and correspondingly the latter is plotted for $\delta \gamma=\gamma_1-\gamma_2=0.07$.}
\label{fig:fermion1}
\end{figure}

We are then left with the determination of the velocities $v_{\psi},v_s$ in addition to $v_b=v_c\cos(\theta_s/2)$, which should be obtained by solving \req{eq:vel1}. Here the trivial solution will do the work:
\begin{equation}
\label{eq:velspin}
	v_{\psi}=v_b=v^\text{BEC}_s=v_c\cos\frac{\theta_s}{2}.
\end{equation}
A comparison of the current semiclassical result and the exact solution is shown in Fig. \ref{fig:fermion1}, they agree well in the vicinity of $P=k_F$.

In between the deep BCS and BEC sides, the physical picture of the $S=1/2$ excitations remain the same - they are moving solitons with one extra fermion bounded on the soliton core. This explains what we observed in exact solutions: instead of adding one particle on the uniform background, the more energy-favorable excitation is the addition of one particle on the dark soliton. The energy cost in the creation of the dark soliton is offset by the energy gain of trapping the particle inside the dip of the density profile. The fact that the minimum energy is achieved exactly at the Fermi momentum is then a consequence of the soliton formation.

\section{Theory of $S=0$ Soliton}
In this section, we apply the general formalism to the $S=0$ soliton, where we will find a crossover between the two weak coupling limits of the soliton structure.

\subsection{Deep BEC Side}
\begin{figure}
\includegraphics[scale=0.5]{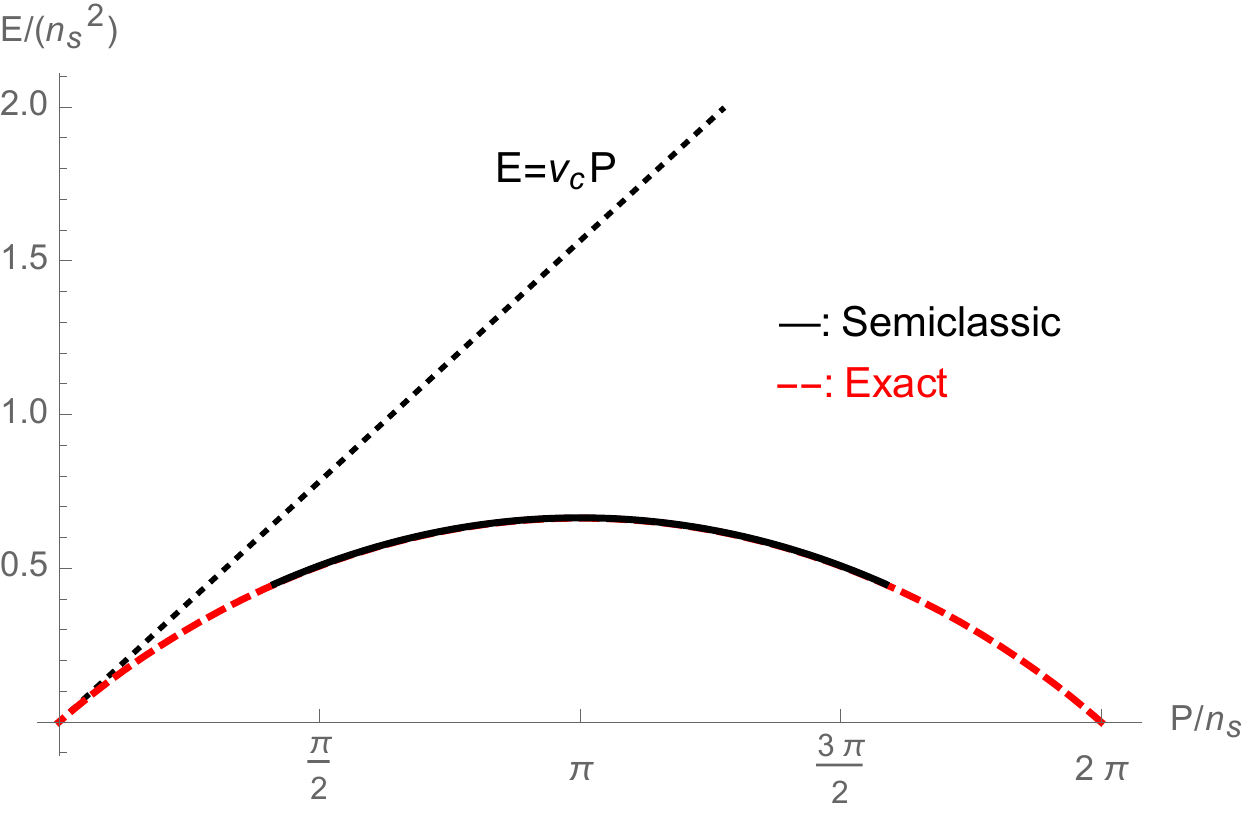}
\caption{\footnotesize The typical $S=0$ excitation spectrum in the semiclassical result and the exact solution, The former is plotted for $\gamma=g/n=0.06$, and correspondingly the latter is plotted for $\delta\gamma=\gamma_1-\gamma_2=0.06$.}
\label{fig:soliton1}
\end{figure}

The analysis on the deep BEC side is simpler, since we have only the Gross-Pitaevskii equation for the classic field $b(z)$ presented in \req{eq:GP}, and we don't need to worry about the self-consistency requirement as in \req{eq:selfcon}. In fact, from \req{eq:GP} we can reconstruct the low energy effective Hamiltonian as
\begin{equation}
\hat{\mathcal{H}}=\int dz\left[\frac{1}{2}\partial_z\hat{b}^{\dagger}\partial_z\hat{b}+g\hat{b}^{\dagger}\hat{b}^{\dagger}\hat{b}\hat{b} \right],
\end{equation}
with is just the Lieb-Liniger model defined in \req{eq:HLL} but with the mass $m_b=1$. As mentioned previously, the fact that $S=0$ (type-II) excitations of the Lieb-Liniger model have the physical interpretation as moving solitons is well understood \cite{Kulish:1976ek,PhysRevA.78.053630}. The energy and momentum can be calculated directly using the soliton profile in \req{eq:Soliton}:
\begin{equation}
\begin{split}
E^\text{BEC}_0(\theta_s)=&\int dz~\left(\frac{1}{2}|\partial_zb|^2+g(|b|^2-n_s)^2\right)\\
=&\frac{4}{3}n_sv_c\sin^3\frac{\theta_s}{2}\\
P^\text{BEC}_0(\theta_s)=&\int dz~\frac{1}{2i}(b^*\partial_zb-b\partial_zb^*)+n_s\theta_s\\
=&n_s(\theta_s-\sin\theta_s),
\end{split}
\end{equation}
then the soliton velocity is determined as $v^\text{BEC}_s=\partial E_0/\partial P_0=v_c\cos(\theta_s/2)$, which is consistent with the result in \req{eq:velspin}. A  comparison of the semiclassical result with the exact solution is shown in Fig. \ref{fig:soliton1}, where in the weak coupling limit, we will obtain a next to perfect match \cite{Kulish:1976ek}.

\subsection{Deep BCS Side}
Now we move on to the deep BCS side, where the situation is complicated by the requirement of the self-consistent condition in \req{eq:selfcon22}. The $S=0$ soliton is obtained by setting $\sum_{\sigma}\lrangle{\hat{\gamma}^{\alpha\dagger}_{+,k,\sigma}\hat{\gamma}^{\alpha}_{+,k,\sigma}}=0$ and $\sum_{\sigma}\lrangle{\hat{\gamma}^{+\dagger}_{0,\sigma}\gamma^{+}_{0,\sigma}}=0$, then \req{eq:selfcon22} reduces to
\begin{equation}
	\Delta=\lambda \int\frac{dk}{2\pi}\frac{\Delta}{\epsilon_k}+\frac{\lambda}{4}\frac{\Delta_0}{v_F}\frac{\theta_s}{\pi}\frac{\sin\frac{\theta_s}{2}}{\cosh^2\left(\frac{\Delta_2}{v_F}z \right)}+\tau(i\partial_z\Delta).
\end{equation}
Compared with \req{eq:selfcon2} for the $S=1/2$ soliton, the self-consistent equation here differs in the second term on the righthand side: it is now proportional to $\theta_s$ instead of $(\theta_s-\pi)$ as in \req{eq:selfcon2}. Since the dark soliton corresponds to the parameterization that $\theta_s=\pi$ and $\tau=0$, this means that we cannot fulfill the self-consistent equation for the $S=0$ soliton thus constructed, and the ground state of $\hat{\mathcal{H}}_{\psi}$ in \req{eq:meanpsi} does not correspond to a proper $S=0$ excitation, as mentioned in section III.

A solution of the above problem was conjectured by ~\citet{PhysRevA.91.023616} and consisted in the assumption that both negative and positive energy localized states are occupied with fractional occupation number. We found this solution to be incorrect for the following reasons: (1) Only positive energy states of the BCS Hamiltonian are meaningful and including the negative energy ones, in fact, describes the same states by different variables; (2) Even if this mistake is rectified, the fractional occupation of the localized state is forbidden in the mean field level as this state is not connected to the continuum (unlike Fano resonance); (3) It gives the value of the energy and of the curvature at $P=k_F$ inconsistent with the exact solution \cite{1367-2630-18-7-075004}.

\begin{figure}[htp!]
	\includegraphics[scale=0.37]{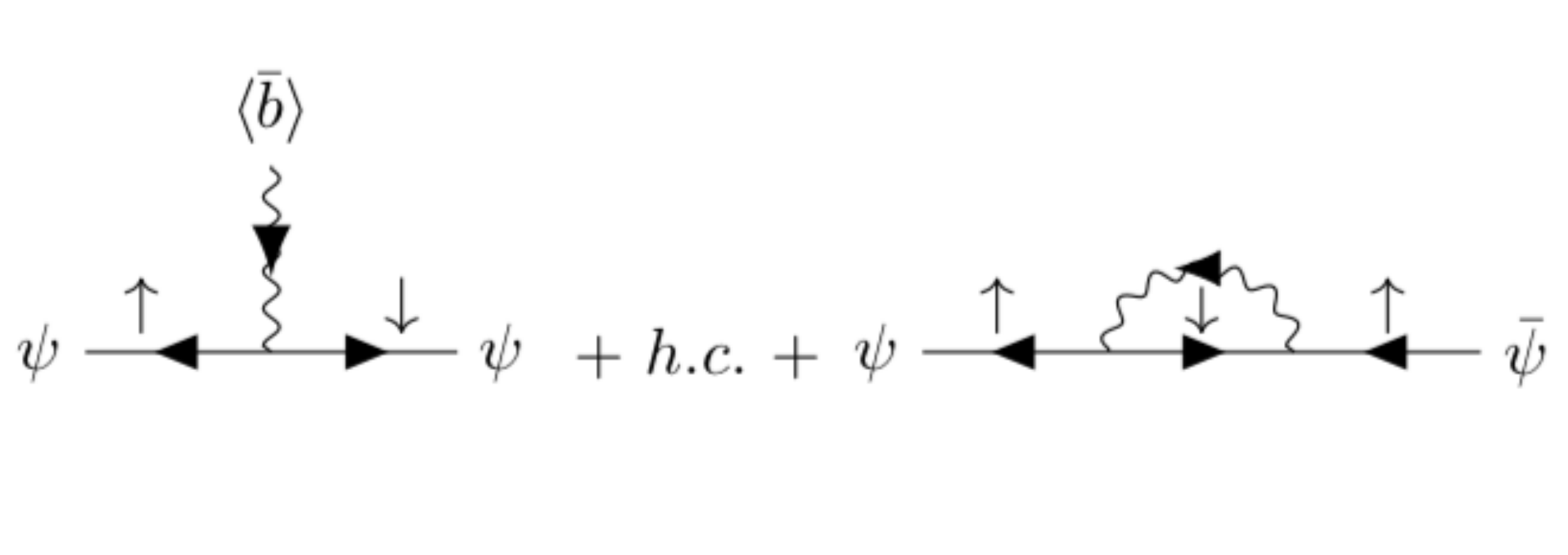}
	\caption{\footnotesize The diagrammatic representation of the mean field potential (left) and the Fock potential (right) experienced by fermions. The solid line denotes the fermion propagator, the wiggled line denotes the boson propagator, and the fermion-boson vertex denotes the resonant coupling $t_b$. The thin arrow on the fermion propagator denotes the spin direction of the fermion.}
	\label{fig:Feynman2}
\end{figure}

Here, inspired by the fact that the maximum energy is on the scale of the Fermi energy, we propose that the proper construction of a $S=0$ soliton is as follows. We break the weakly bounded pair at the bottom of the Fermi sea, which leaves us with two fermions. We then put one of them on the localized level to produce a $S=1/2$ soliton. This is possible because the breaking of the bound pair at the bottom of the Fermi sea has no effect on the linearized spectrum. After that, we can form a singlet from the other fermion and the $S=1/2$ soliton, which gives us the desired $S=0$ soliton. To carry out such a construction, we need to go beyond the present mean field analysis and include the Fock potential produced by the spin density on fermion of the opposite spin (see Fig. \ref{fig:Feynman2}). Hartree potential is not considered here since it is not sensitive to spin.

\begin{figure}[htp!]
	\includegraphics[scale=0.24]{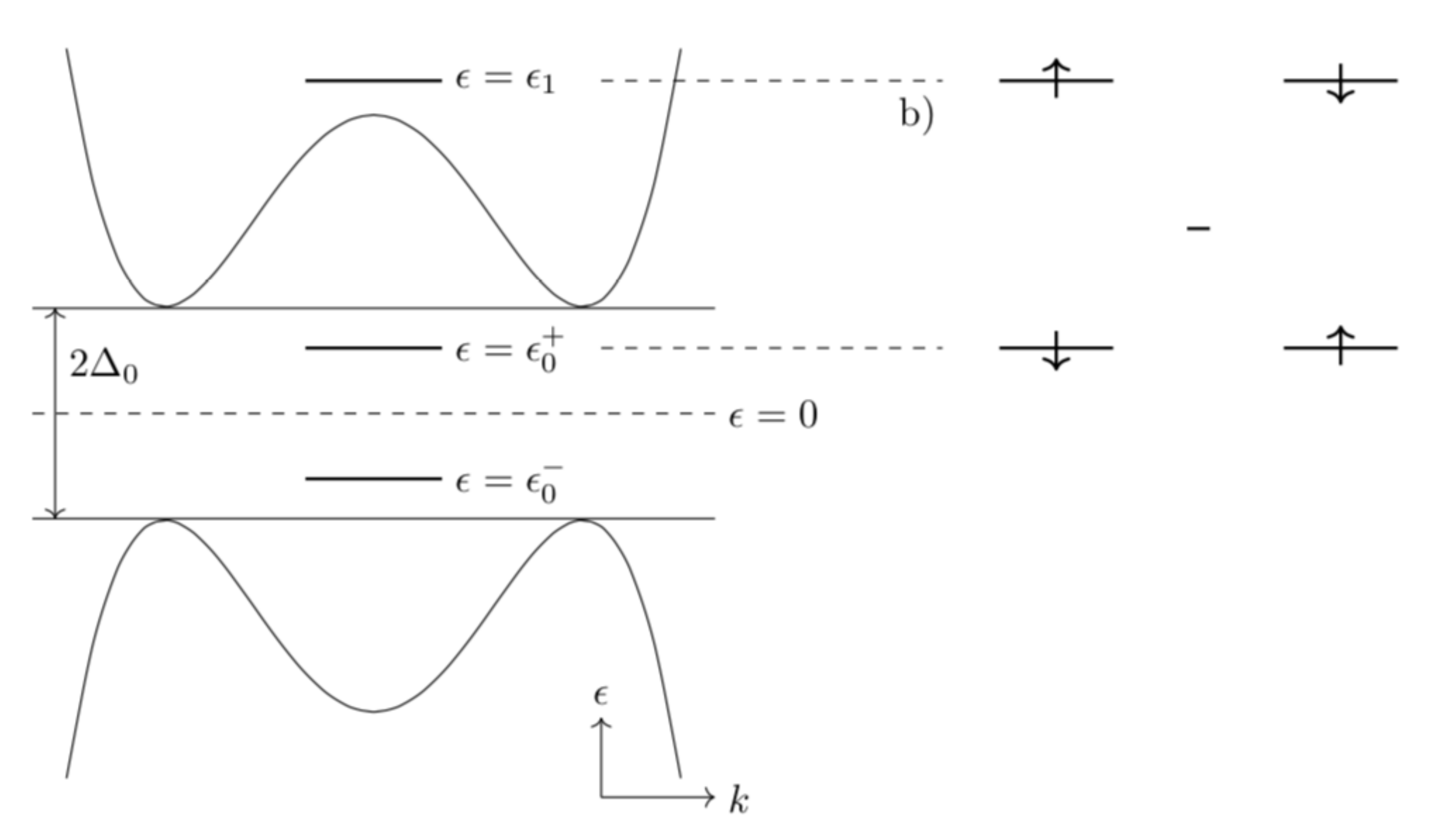}
	\caption{\footnotesize a) The spectrum in the Fock approximation, where $\epsilon^{\pm}_0$ represents the two localized states in \req{eq:localeigen}. They are linear combinations of the left and right moving localized states in \req{eq:local1} and \req{eq:local2} once nonlinear effects are taken into consideration. $\epsilon_1$ represents the extra localized state produced by the Fock potential. b) The configuration of the $S=0$ soliton, which is formed as a singlet of the two localized states with energies $\epsilon_1$ and $\epsilon^{+}_0$.}
	\label{fig:filling}
\end{figure}

By including the Fock potential, the equation of motion for the fermionic fields is modified as:
\begin{equation}
\label{eq:modeequv}
\begin{split}
&\begin{pmatrix}-i\alpha v_F\partial_z-\alpha v_{\psi}k_F & \Delta(z)\\ \Delta^*(z) & i\alpha v_F\partial_z-\alpha v_{\psi}k_F \end{pmatrix}\begin{pmatrix}u^{\alpha}_n\\ v^{\alpha}_n\end{pmatrix}\\
&+\begin{pmatrix}-V_{\text{F}} & 0\\ 0 & V_{\text{F}}\end{pmatrix}\begin{pmatrix}u^{\alpha}_n\\ v^{\alpha}_n\end{pmatrix}=\bar{\epsilon}^{\alpha}_n\begin{pmatrix}u^{\alpha}_n\\ v^{\alpha}_n\end{pmatrix}.
\end{split}
\end{equation}
The Fock potential is $V_{\text{F}}=\frac{\lambda}{2}\frac{\Delta_2}{v_F}\text{sech}^2\left( \frac{\Delta_2}{v_F} \right)$, where we have incorporated the constant part $\lambda n/2$ of $V_F$ into the chemical potential. Also from \req{eq:vbbb2} we have $v_{\psi}\approx v^{\text{BCS}}_s$. The first term in $V_{\text{F}}$ comes from the fermions in the continuous spectrum and the second term comes from the fermion in the localized state. For states with momentum near $k_F$, the Fock potential $V_{\text{F}}$ only acts as a small correction to the chemical potential, while for states near zero momentum, $V_{\text{F}}$ has a more dramatic effect of producing an extra localized state. In the latter case, we can ignore the small off-diagonal components in \req{eq:modeequv}, and the hole excitation near zero momentum is described by the Schrodinger equation without linearization:
\begin{equation}
\label{eq:linearconfine}
\left(\partial^2_z+\mu+iv_{\psi}\partial_z\right)\psi(z)+V_{\text{F}}\psi(z)=\bar{\epsilon}\psi(z).
\end{equation}
As usual, we perform the gauge transformation that $\psi(z)\to \psi(z)e^{-iv_{\psi}z/2}$ with a shift in momentum as $-v_{\psi}/2$, and again we are led to the Schrodinger equation for a particle moving in P$\ddot{\text{o}}$schl-Teller potential:
\begin{equation}
\label{eq:ExtraBound}
-\partial^2_z\psi-\alpha^2\zeta(\zeta-1)\text{sech}^2\left( \alpha z \right)\psi=\left(-\bar{\epsilon}+\mu +\frac{v_{\psi}^2}{4}\right)\psi,
\end{equation}
where $\alpha=\frac{\Delta_2}{v_F}$, $\alpha^2\zeta(\zeta-1)=\frac{\lambda\Delta_2}{2v_F}$. This produces a bound hole state with energy
\begin{equation}
\label{eq:ExtraBound3}
\bar{\epsilon}_1=\frac{\Delta_0^3}{32\lambda v^3_F}\left(1-\frac{3}{8}\xi^2\right)+\mu+\frac{v^2_{\psi}}{4}.
\end{equation}
Also the momentum of this bound hole state is simply $k_1=-v_{\psi}/2$, and we can determine the eigenvalue $\epsilon_1$ that contributes to the energy $E$ as
\begin{equation}
	\epsilon_1=\bar{\epsilon}_1+v_{\psi}k_1.
\end{equation}
This localized state then combines with the $S=1/2$ soliton to form a singlet, which is the desired $S=0$ soliton (see Fig. \ref{fig:filling}).

\begin{figure}
\includegraphics[scale=0.5]{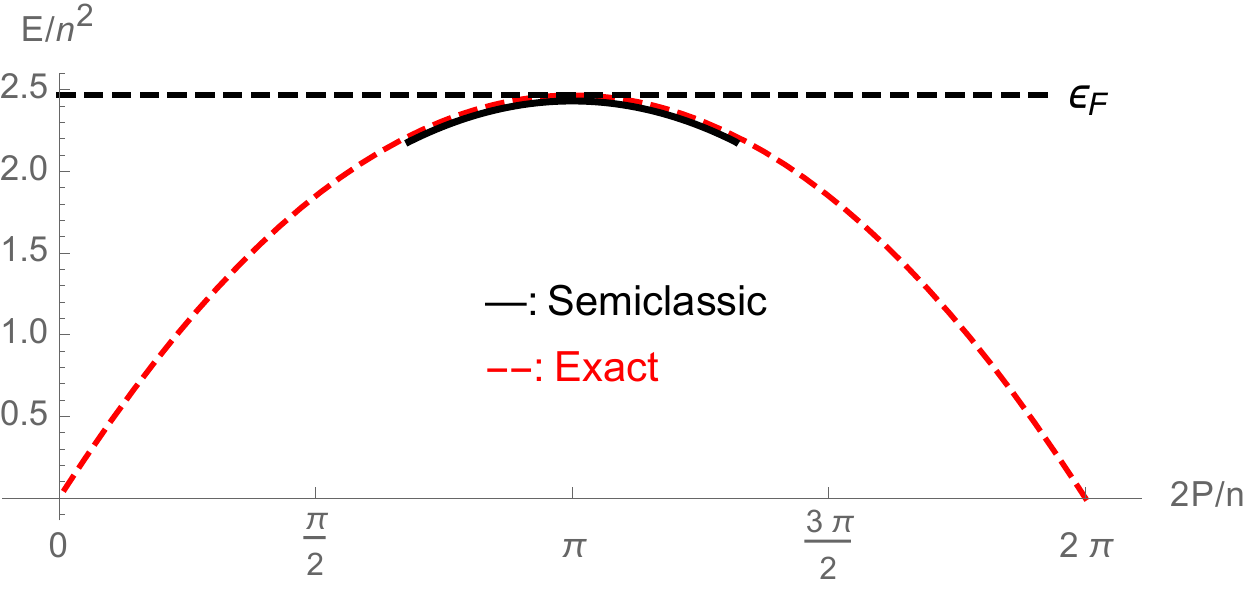}
\caption{\footnotesize The typical $S=0$ excitation spectrum in the semiclassical result and the exact solution. The latter is plotted for $\gamma=c_F/n=0.15$, and correspondingly the former is plotted taking the spin gap at the same coupling strength as the input parameter.}
\label{fig:soliton2}
\end{figure}

We can now determine the energy and the momentum of the $S=0$ soliton as
\begin{equation}
\begin{split}
E^\text{BCS}_0(\theta_s)& =E^\text{BCS}_{1/2}(\theta_s)+\epsilon_1\\
&= E_0+\left(\frac{\Delta_0}{4\pi}-\frac{3}{8}\frac{\Delta_0^3}{32\lambda v^3_F}-\frac{v_F^2}{4\pi^2}\right)\xi^2,\\
& E_0=\frac{2\Delta_0}{\pi}+\mu+\frac{\Delta^3_0}{32\lambda v_F^3},\\
P^\text{BCS}_0(\theta_s)&=P^\text{BCS}_{1/2}(\theta_s)+k_1 \\
	& =k_F+\left(\frac{v_F}{2\pi}-\frac{\Delta_0}{2v_F}\right)\xi
\end{split}
\end{equation}
where we have used the expression for $v_{\psi}\approx v^{\text{BCS}}_{s}$ in \req{eq:velocity}. In the weak coupling limit, we have $v^2_F\gg \Delta_0$, then the energy does conform to what we observed in exact solutions that it is on the scale of the Fermi energy
$\mu=\epsilon_F$, and the dispersion of the $S=0$ soliton can be approximated as
\begin{equation}
\begin{split}
E_0(P_0)\approx E_0-(P_0-k_F)^2.
\end{split}
\end{equation}
which agrees with the exact solutions and reduces to the noninteracting fermion result. A comparison of the current semiclassical result with the exact solution is shown in Fig. \ref{fig:soliton2}, where the former grasps the basic features of the latter.

\subsection{Crossover Problem}
Here we argue that the crossover region of the $S=0$ soliton is not described by a simple mean field configuration but rather by the linear combination of the states considered in subsections A and B.

Unlike the $S=1/2$ soliton, the $S=0$ soliton on the BEC side and BCS side have different natures. The former is just the usual soliton formed in the condensed bound pairs, while the latter is a singlet formed by two localized spins (one is trapped by the Fock potential of the other). We refer to the latter as a dressed soliton. The dressed soliton can tunnel into the usual soliton configuration since the state localized by the Fock potential lies in the continuous spectrum (see Fig. \ref{fig:filling}). On the deep BCS side, the tunneling is negligible. When we tune the resonant level to leave the deep BCS side, the tunneling between the dressed soliton and the usual soliton becomes stronger, and the physical soliton will be a linear combination of them. Till on the deep BEC side, the usual soliton dominates. The two localized spins we have on the BCS side then bound together to become one of the bound pairs on the BEC side. There is no abrupt change happening in the soliton formation along the crossover, just as what we have observed in the excitation spectra of the exactly solvable models.

The above qualitative argument can be made more rigorous by analyzing the tunneling of the state localized by the Fock potential into the quasiparticle continuum. The desired analysis is performed for \req{eq:modeequv} in the regime where the chemical potential $\mu$ is the largest energy scale near the BCS side, so the off-diagonal part can be treated perturbatively. We have both electron-like eigenstate $\ket{\Psi^e}$ and hole-like eigenstates $\ket{\Psi^h}$ at zeroth-order, and in each sector we will get a localized state $\ket{\Psi^{e,h}_0}$. Here we focus on the state $\ket{\Psi^h_0}$ with energy $\epsilon^h_0$ on the scale of $\mu$, which will tunnel into the continuum of the electron-like state $\ket{\Psi^e_k}$ as the off-diagonal perturbation sets in. The resonance width due to this tunneling can be calculated using the Fermi golden rule:
\begin{equation}
	\Gamma=2\pi \nu_{\uparrow,\downarrow}(2\epsilon_F)|\mathcal{M}|^2=\frac{1}{\sqrt{2}v_F}|\mathcal{M}|^2,
\end{equation}
where we have taken the density of states $\nu_{\uparrow,\downarrow}(\epsilon)=\nu(\epsilon)/2$ for each spin to be the one at $2\epsilon_F$ since the localized level is close to the chemical potential $\mu$, and $\mathcal{M}=\bra{\Psi^e_k}\begin{pmatrix}0 & \Delta \\ \Delta^* & 0 \end{pmatrix}\ket{\Psi^h_0}$ is the matrix element between the continuum and the localized state. It can be estimated by the Fourier component $\tilde{\Delta}(\sqrt{2}k_F)$ of the soliton profile, normalized by the size of the soliton:
\begin{equation}
\label{eq:Gamma}
	|\mathcal{M}|^2=\frac{1}{l_s}|\tilde{\Delta}(\sqrt{2}k_F)|^2=\frac{\Delta^2_0l_s\pi^2\sin^2\frac{\theta_s}{2}}{\sinh^2\left( \frac{l_s\pi k_F}{\sqrt{2}} \right)}
\end{equation}
In the end we obtain the following result for the resonance width near the BCS side:
\begin{equation}
	\Gamma=\frac{\Delta^2_0l_s\pi^2\sin^2\frac{\theta_s}{2}}{\sqrt{2}v_F\sinh^2\left( \frac{l_s\pi k_F}{\sqrt{2}} \right)}
\end{equation}
For the large chemical potential near the BCS side, the ratio between the resonance width $\Gamma$ and the energy $\epsilon^h_0$ is exponentially small and the localized state remains well-defined. This is equivalent to the statement that the tunneling to the usual soliton is negligible. As we tune the system away from the BCS side, up to the point where the chemical potential is comparable to the gap parameter $\Delta_0$, the velocity of the particle and the Fermi momentum are also tuned to be on the order of magnitude comparable to $\Delta_0$. At that point, the resonance width $\Gamma$ becomes comparable to the energy $\epsilon^h_0$. With further tuning toward the BEC side, we then encounter a large resonance width $\Gamma\gg \epsilon^h_0$, and the localized state ceases to be well defined and merges into the quasiparticle continuum. Correspondingly, we have the spin-singlet described on the BCS side develop into a normal bound pair on the BEC side. As a result, we have a smooth crossover from the soliton of BCS type into the one of BEC type, while the mathematical description of the state obtained from the tunneling between different mean field solutions is beyond the scope of this paper.

\section{Conclusion}
In this paper, we developed a semiclassical theory of moving solitons in one dimensional BCS-BEC crossover, where on both the deep BCS and deep BEC side, our results grasp the essential features of the exact solution. Our theory also resolves the inconsistency between the semiclassical analysis and the exact solutions in the attractive Yang-Gaudin model. In the meantime, we revealed the mechanism of a striking phenomenon discussed in our previous paper that the minimum energy of the spin excitation is fixed at the Fermi momentum along the whole range of BCS-BEC crossover in one dimension. Conventionally in higher dimensions, we would expect this momentum to be shifted from $k_F$ on the BCS side to zero somewhere on the way to the BEC side, and it is believed that this is the only sharp change that could happen in a BCS-BEC crossover \cite{2015qgee.book..179P}. We then show that the counterintuitive fixing comes about as a special feature of the one dimensional systems, that the conventional quasiparticle is not stable with respect to soliton formation. Our theory serves as yet another example of the important role solitons can play in low dimensional physical systems, in addition to those well established in one dimensional lattice models \cite{PhysRevB.22.2099,PhysRevB.21.2388} and charge density waves \cite{Brazovskii_1989,1980JETP...51..342B}.

\section{Acknowledgment}
We would like to thank Victor Gurarie for helpful discussions and valuable comments on the paper. This work is supported by Simons Foundation.

\bibliography{Soliton}

\end{document}